\documentclass[aoas]{imsart}

\RequirePackage{amsthm,amsmath,amsfonts,amssymb}
\usepackage{bbm}
\RequirePackage[authoryear]{natbib}
\RequirePackage[colorlinks,citecolor=blue,urlcolor=blue]{hyperref}
\RequirePackage{graphicx}
\RequirePackage[ruled,vlined]{algorithm2e}
\RequirePackage{algpseudocode}
\RequirePackage{float}
\usepackage{subcaption}
\usepackage{booktabs} 
\usepackage{tabularx}
\usepackage{ulem}
\usepackage{enumitem}

\usepackage[left]{lineno}

\startlocaldefs
\theoremstyle{plain}

\theoremstyle{remark}

\DeclareMathOperator*{\argmax}{arg\,max}
\DeclareMathOperator*{\argmin}{arg\,min}
\DeclareMathOperator{\E}{\mathbb{E}}

\endlocaldefs

\begin{document}

\begin{frontmatter}
\title{A time warping model for seasonal data with application to age estimation from narwhal tusks}
\runtitle{a time warping model}

\begin{aug}
\author[A]{\fnms{Lars N.}~\snm{Reiter}\ead[label=e6]{lrn@math.ku.dk} \orcid{0000-0002-0931-3762}},
\author[A]{\fnms{Adam G.}~\snm{Hoffmann}\ead[label=e1]{adam.hoffmann@sund.ku.dk} \orcid{0009-0008-0058-3804}},
\author[B]{\fnms{Mads Peter}~\snm{Heide-Jørgensen}\ead[label=e4]{mhj@mail.ghsdk.dk} \orcid{0000-0003-4846-7622}},
\author[B]{\fnms{Eva}~\snm{Garde}\ead[label=e3]{evga@mail.ghsdk.dk} \orcid{0000-0003-3016-5722}},
\author[C]{\fnms{Adeline}~\snm{Samson}\ead[label=e2]{adeline.leclercq-samson@univ-grenoble-alpes.fr}},
\author[A]{\fnms{Susanne}~\snm{Ditlevsen}\ead[label=e5]{susanne@math.ku.dk} \orcid{0000-0002-1998-2783}}
\address[A]{Department of Mathematical Sciences, University of Copenhagen, Copenhagen, Denmark \printead[presep={,\ }]{e6,e5,e1}}
\address[B]{Greenland Institute of Natural Resources \printead[presep={,\ }]{e3,e4}}
\address[C]{ Jean Kuntzmann Laboratory, Université Grenoble Alpes \printead[presep={,\ }]{e2}}
\end{aug}

\begin{abstract}
Signals with varying periodicity frequently appear in real-world phenomena, necessitating the development of efficient modelling techniques to map the measured nonlinear timeline to linear time. Here we propose a regression model that allows for a representation of periodic and dynamic patterns observed in time series data. The model incorporates a hidden strictly positive stochastic process that represents the instantaneous frequency, allowing the model to adapt and accurately capture varying time scales. A case study focusing on age estimation of narwhal tusks is presented, where cyclic element signals associated with annual growth layer groups are analyzed. We apply the methodology to data from one such tusk collected in West Greenland and use the fitted model to estimate the age of the narwhal. The proposed method is validated using simulated signals with known cycle counts and practical considerations and modelling challenges are discussed in detail. This research contributes to the field of time series analysis, providing a tool and valuable insights for understanding and modeling complex cyclic patterns in diverse domains.
\end{abstract}


\end{frontmatter}

\section{Introduction} \label{sec:intro}

Time series data with cyclic patterns are commonly encountered in real-world applications, ranging from environmental, ecological, physical and physiological studies to economic forecasting \citep{GlassMackey, ShumwayStoffer}.
Understanding and accurately modelling such patterns are crucial for gaining insights into underlying dynamics and making reliable conclusions. 

In some applications, the measurements are taken along a segment, where each measured point at some distance corresponds to an unknown time. Key examples are tree rings, sea sediment cores, ice cores, hair, ear stones \citep{46,47,48,49,50}, or in this case, narwhal tusks. A given variable is measured along a transect line at equidistant spatial points, and the oscillation patterns in the data are hypothesized to reflect yearly variations. The measurements can then be dated, i.e., distances can be transformed to a timeline. However, manual counting of cycles is often highly uncertain, and typically varies both between and within experts. 

A significant challenge presented by such datasets relates to the unknown nonlinear relation between distance and the underlying timeline, complicating the precise mapping of the signals to their actual temporal positions. In cases where reference signals are available or when the primary objective of the study centers on signal alignment rather than precise temporal mapping, methodologies like Graphical and Dynamic Time Warping \citep{35,36,37} offer effective solutions for tasks involving pattern matching and classification. However, when reference signals are absent, and the exact mapping from spatial position to real time is of particular concern, the literature on time warping frameworks targeting cyclic signals is limited. One example is \cite{61}, who developed a particle-EM variant that recovers the hidden phase and cycle shape jointly. Another example is \cite{62}, who proposes two ways to estimate the time warping function in signals with irregular cycles, either directly estimating the warping functions or by using a probabilistic approach to construct a suitable estimator. 

This paper presents a novel approach for modelling noisy non-stationary cyclic signals with changing periodicity and amplitude. As opposed to the traditional seasonal decomposition \citep{11,12} where a signal is split into trend, seasonal, cyclic, and noise components, our model incorporates the seasonal and cyclic effects, and combine them into a single component. A key innovation in our model is the incorporation of a hidden strictly positive stochastic process, specifically the square-root diffusion process \citep{soerensen2012,DitlevsenLansky2006}, also known as the Cox-Ingersoll-Ross (CIR) process in the econometrics literature \citep{55}. 
The integral of this process is strictly increasing and models the material growth, by modulating the frequency while also respecting the direction of time. This provides realism in capturing the varying periodicity observed in real-world data. 

Estimation within our model is achieved using a stochastic variant of the EM algorithm, where we repeatedly filter the unobserved square-root diffusion process using Sequential Monte Carlo (SMC) and improve our estimates of the parameters in each new run. This procedure is explained in detail in section \ref{sec:estimation}. While our model has applications across various domains, we focus here on its use in an ecological context. In this field, time series data are commonly associated with annual cycles, such as the growth patterns of organisms where various exogenous and endogenous factors can influence the speed of growth \citep{13,14,15}. Our method can effectively capture the changing frequencies and provide more accurate predictions for these cyclic patterns. 


We illustrate the method with data obtained from a narwhal (\textit{Monodon monoceros}) tusk. The narwhal is a medium-sized odontocete (toothed whale), endemic to the Arctic regions.
Understanding the age structure of wildlife populations is critical for assessments and conservation efforts \citep{watt2020estimating}. Age estimates form the basis of age-structured population models, which are essential tools for predicting demographic trends and informing management strategies, particularly for harvested or vulnerable species \citep{garde2022biological}. Despite its importance, reliable age estimation in narwhals remains a challenge, and no fully accepted standard method has yet been established \citep{read2018review}.

Narwhals are best known for the elongated, spiraled tusk of the male. Female narwhals rarely develop tusks; instead, they possess two small, embedded maxillary teeth. In males, the left maxillary tooth begins to erupt around 1–2 years of age and gradually elongates into the long tusk, which continues to grow throughout life. In contrast, the embedded teeth in both sexes generally cease growing between the ages of 10 and 20 years \citep{zhao2021ontogenetic,watt2020estimating}. 

Counting annual growth layer groups (GLGs)—each consisting of an opaque and a translucent layer—in dental tissues has traditionally been used to estimate the age of mammals \citep{hay1980age, read2018review}. However, this method poses several challenges in narwhals \citep{hay1980age, zhao2021ontogenetic}. For the embedded tooth, dentinal occlusion at sexual maturation obscures GLGs resulting in a minimum age estimate. For the erupted tusk,  natural wear can gradually erode the tip of the tusk, removing early GLGs. This is mostly observed in older individuals, but will, as with the embedded tooth, result in a minimum age estimate. In addition, tusks seldom erupt in female narwhals and their large size, and the logistical and financial costs of collecting tusks limit its applicability in large-scale studies \citep{Garde2012}. 

The most significant challenge, however, lies in the nature of the GLGs themselves: they are often faint, compressed or with subtle transitions between layers, making manual counting difficult. This process requires an experienced specialist, is inherently subjective and non-reproducible, and is associated with high variability both within and between observers.

Researchers have therefore explored two alternative techniques for age estimation. One promising approach is aspartic acid racemization (AAR), which has shown potential for estimating age in narwhals and other marine mammals \citep{garde2007age, garde2015life,watt2020estimating}. However, the initial development and calibration of the AAR method in narwhals relied on manual GLG counts from tusks \citep{Garde2012}, introducing uncertainty due to the inherent subjectivity of GLG interpretation. Additionally, AAR analysis requires access to the eye lens at the time of death, which can be logistically challenging to obtain.
A recent study \cite{Gardeetal2024} used radiocarbon ($^{14}$C) dating to assess the accuracy of manual GLG-based age estimates in three tusks previously analyzed in \cite{Garde2012}. This method leverages the so-called "bomb pulse" - a sharp increase in atmospheric radiocarbon resulting from nuclear weapons testing in the mid- to late 1950s—which rapidly spread through global food webs, including marine ecosystems in the North Atlantic and Arctic Oceans. The bomb pulse serves as a temporal marker, allowing researchers to distinguish between pre- and post-pulse deposition of radiocarbon in narwhal tusks. However, this method is only of value for setting a benchmark for the bomb pulse and it can only be applied to whales born before or around the time of the bomb pulse.
Radiocarbon dating results have shown that the manual GLG counts may overestimate the possible age. This is likely due to the challenges of reading tusks from older animals, for the reasons mentioned earlier.
Our proposed model has the potential to reduce the subjectivity associated with manual GLG counting and to enhance our understanding of tusk growth in narwhals, such as the element composition in narwhal tusks , which may reflect underlying biological or environmental changes. In addition to its application in age estimation of narwhals and other toothed marine mammals, this method holds potential for broader use in any biological material that exhibits cyclic growth patterns.

In Section \ref{sec:model}, we outline the model, in Section \ref{sec:estimation} we propose how to estimate model parameters and in Section \ref{sec:simulation} we perform a simulation study. In Section \ref{sec:case} we demonstrate the model and estimation technique in a narwhal tusk, where we focus on the mineralized elements (specifically the isotope Barium-137), which are deposited in the same annual pattern as the growth layers. For these data, decoding the hidden stochastic process of the model corresponds to uncovering the process that drives the growth of the tusk. Unlocking the growth process is of scientific interest, because it reveals life chapters of accelerating or declining growth, which may reflect underlying biological or environmental changes \citep{60}. From this process, we are then able to estimate the number of cycles, thereby also the age of the narwhal. We conclude with a discussion of the model in section \ref{sec:disc}.

\section{Time warping model}\label{sec:model}
Consider a regression model with additive noise, 
\begin{equation} 
y_i =  f(x_i, \theta) + \epsilon_i, \quad i = 0,1, \ldots , n, \, x_i = i \Delta,  \label{eq:1} 
\end{equation}
where $y_i$ is the measured variable (e.g., a chemical element) at spatial location $x_i$ and $\Delta = x_{i+1}-x_i$ is the distance between measurements. The observation interval is thus $[0, n\Delta]$. The function $f$ is detailed below and depends on parameters $\theta$. The error terms $\epsilon_i$ are assumed independent and normally distributed with mean 0 and variance $\sigma^2$. The variable $x$ is the distance along a cross section of some object of interest, where a specific position $x_i$  corresponds to an unknown time through a monotonic and possibly non-linear function $g$,
\begin{equation}
t_i = g(x_i).
\end{equation}
We impose that $g(x)$ is a strictly increasing function with respect to $x$, to ensure that time does not go backward.   

\begin{figure}[!tb]
    \centering
    \includegraphics[width=.8\linewidth, trim={2cm 1cm 1cm 2cm}, clip]{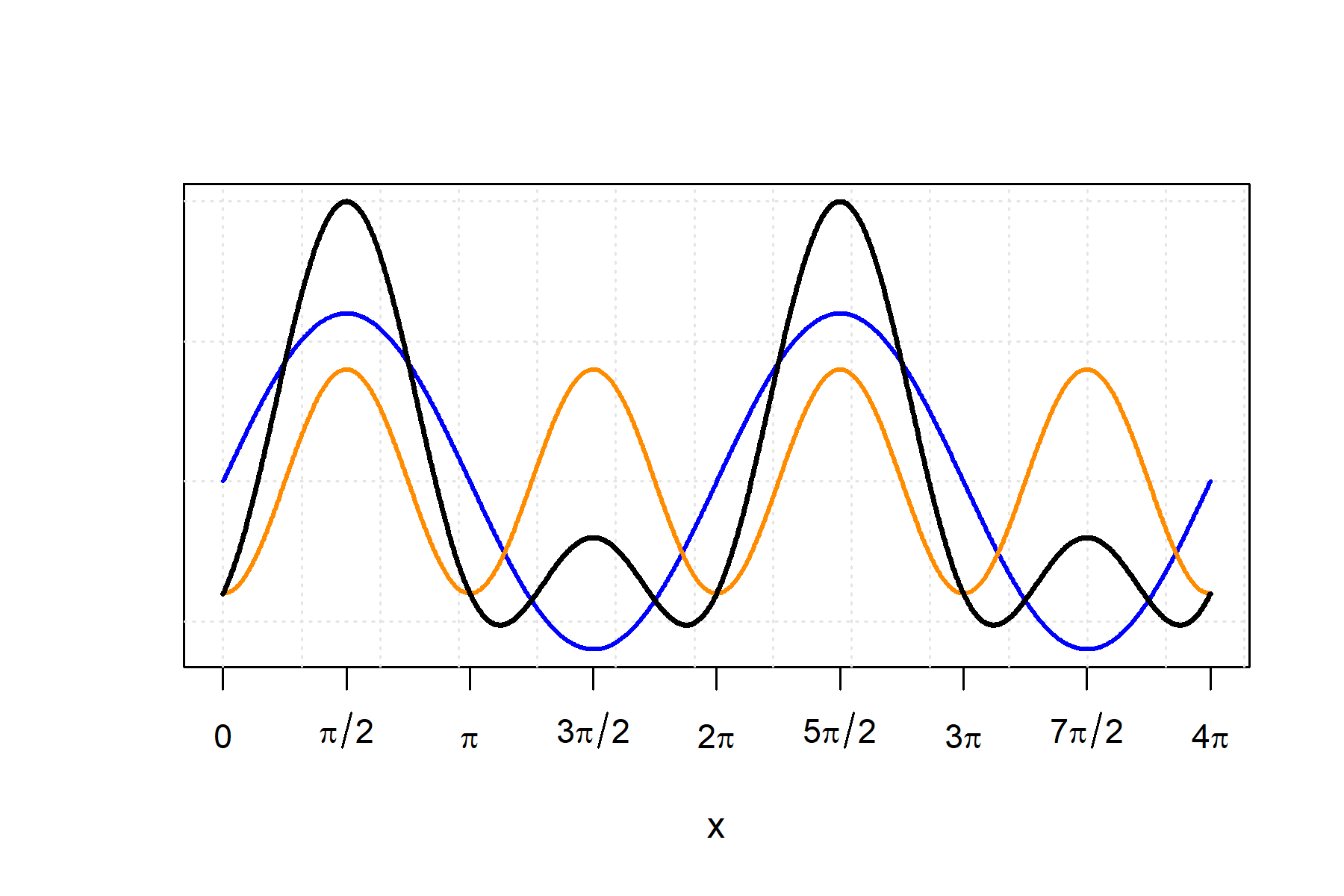}
    \caption{\textbf{Signal composition.} Slow sine wave (blue) and fast cosine wave (orange) as well as their composition $f(x ; \theta)$ (black), using $A=0.6$, $B = 0.4$ and $b=0$ in Eq. \eqref{eq:1b}. The growth process is non-stochastic and linear, $g(x) = x$.}
    \label{fig:harm}
\end{figure}

For the signal, we assume linear combinations of sines and cosines to obtain oscillatory behavior. In particular, for the narwhal tusks, the data show clear indications of a large amplitude slow oscillation intertwined with lower amplitude oscillations of double frequency, with skewed peaks towards higher values. We therefore assume the form
\begin{equation} f(x ; \theta)  =  A \sin(g(x)+b) - B\cos(2g(x)+2b), \quad g(0) = 0,
 \label{eq:1b}
\end{equation} 
where $A > 0$ and $B > 0$ determine the amplitudes and $b$ is the phase at $x=0$. We require $A>B$ for identifiability, and also that $A/B$ is not too small - see supplementary material for a detailed discussion of both criteria. The signal \eqref{eq:1b} is composed of a low-frequency sine wave and a fast-frequency cosine wave: it traverses two cycles for each cycle of the sine wave. The sine wave may, for instance, represent annual cycles, while the fast cosine wave captures the summer-winter variations. See Figure \ref{fig:harm} for an illustration. A generalization to higher frequency terms, for example, a model with higher order harmonics, is discussed in supplementary material. 

There is no trend included in model \eqref{eq:1}, and we assume that it is removed in a prepossessing step, if necessary. For the signals used in our case study, no apparent trend was detected and only the average was subtracted. Furthermore, model \eqref{eq:1b} assumes constant amplitudes. We address this in Section \ref{ssec:preproc}.

In the linear and deterministic case where $g(x) = ax$ the cycles occur regularly (Figure \ref{fig:harm}). 
This is unrealistic in most settings, where the length of one cycle will typically vary due to environmental factors, leading to quasi-periodic dynamics. To describe this, we allow the instantaneous growth rate to be stochastic. We denote the unknown stochastic transformation $g(x)$ as \textit{the growth process}.
Let $g(x) = \int_{0}^x \xi_s d s$ where $(\xi_x)_{x \geq 0}$ is a positive stochastic process. For $(\xi_x)_{x \geq 0}$, we propose the square root diffusion process, satisfying the stochastic differential equation (SDE)
\begin{equation}d \xi_x = -\beta(\xi_x-a)dx + \omega \sqrt{\xi_x}dW_x, \quad \xi_0 = 0,
\label{eq:2} 
\end{equation}
where $W_x$ is a standard Wiener process. The hidden process $(\xi_x)_{x \geq 0}$ describes instantaneous frequency deviations from the baseline frequency $a/2\pi$. The parameter $\beta$ determines the rate of adjustment to the baseline and $\omega$ scales the noise level. This is an ergodic process with stationary distribution the Gamma distribution with shape parameter $2\beta a /\omega^2$, scale parameter $\omega^2/(2\beta)$, mean $a$ and variance $a\omega^2/2\beta$, provided that $\beta, a, \omega > 0$, and $2\beta a \geq \omega^2$ \citep{DitlevsenLansky2020}. These parameter restrictions ensure that the stochastic process generated by Eq. \eqref{eq:2} is strictly positive with probability one (except at $x=0$) with long-term mean $a$, leading to a strictly increasing process $g(x)$.
Denote the one-lag autocorrelation of the growth process by $\rho = \exp(-\Delta \beta)$ and the stationary variance by $\gamma^2 = a \omega^2/2\beta$.
The transition density, i.e., the distribution of $\xi_{x+ \Delta}$ given $\xi_{x}$, follows a non-central $\chi^2$ distribution scaled by a constant factor. Specifically, $\xi_{x+\Delta} | \xi_{x} \sim Z_{x}/2c$, where $Z_{x} \sim \chi^2_{\nu}(\lambda_{x})$ is non-central chi-squared distributed with $\nu = 4a\beta/\omega^2$ degrees of freedom, non-centrality parameter $\lambda_{x+\Delta} = 2c\rho\xi_{x}$ and $c = 2\beta/(1-\rho)\omega^2$ \citep{DitlevsenLansky2006}.

Our main goal is to infer the growth process $g(x)$ in order to date the observations $y_i$ to $t_i = g(x_i)$, $i = 0, \ldots, n$. From $g(x)$ we obtain the number of cycles in the signal: \begin{equation}
    \label{eq:numberofcycles}
    \textrm{Number of Cycles} = \frac{g(x_n)}{2\pi}.
\end{equation}



\begin{figure}[!tb]
    \centering
    \begin{minipage}{0.5\linewidth}
        \centering
        \includegraphics[width=\linewidth]{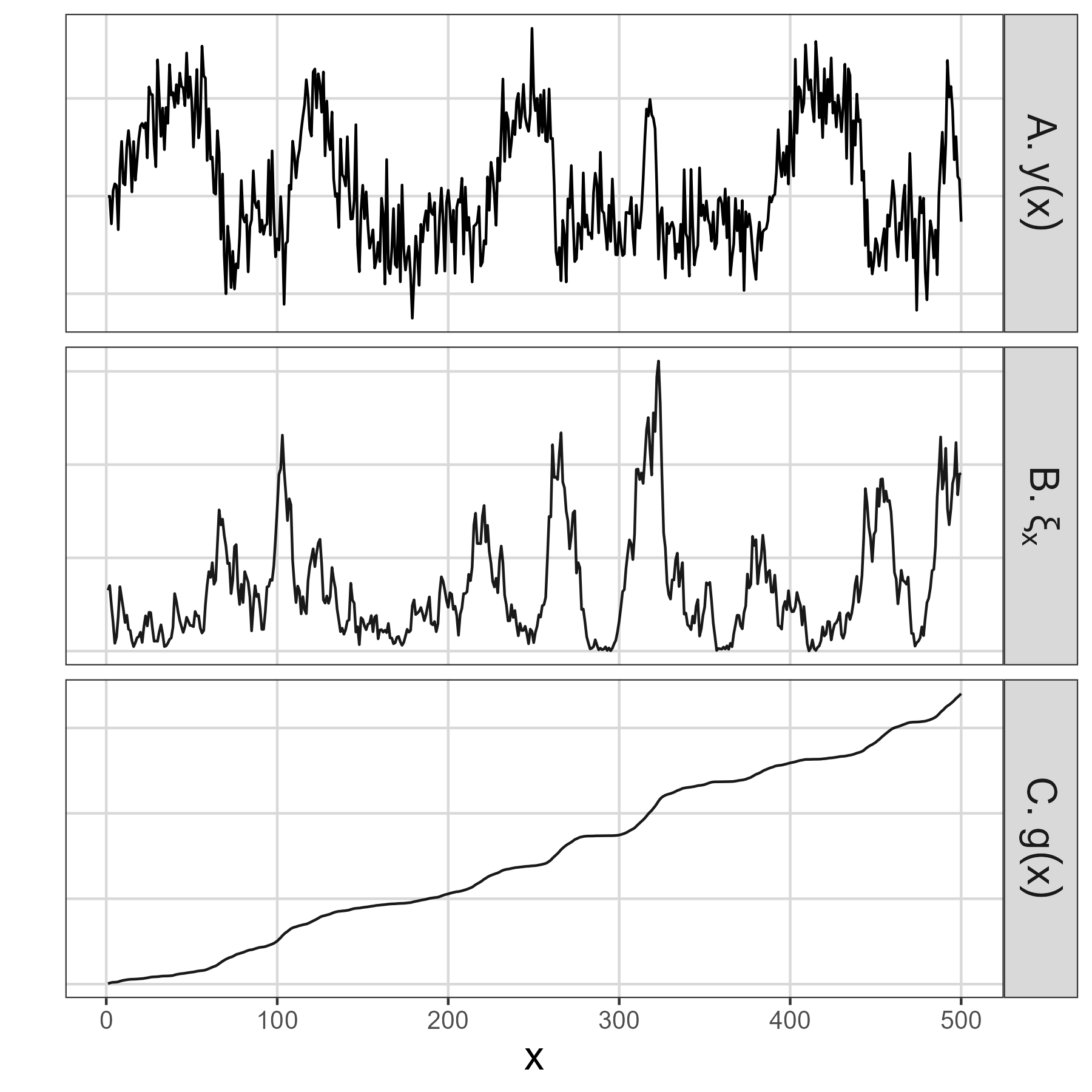}
    \end{minipage}%
    \hfill
    \begin{minipage}{0.45\linewidth}
        \caption{{\bf Example of a simulated signal}. The dataset was simulated using $n=500$, $\Delta = 1$, $\beta = 0.07$, $\sigma = 0.3$,  $A = 0.6$, $B = 0.4$, $a=0.05$, $b = \pi/20$, $\omega^2 = 0.064$. A. Simulated signal with measurement noise $y(x)$. B. Simulated hidden process $(\xi_x)_{x \geq 0}$. C. Simulated growth process $g(x)$.}
        \label{fig:SimSignal}
    \end{minipage}
\end{figure}

The parameters of the model are summarized in Table \ref{tab:pars}. In Figure \ref{fig:SimSignal} we illustrate a signal simulated from model \eqref{eq:1} and \eqref{eq:1b}, along with the embedded growth process $g(x)$ and the instantaneous frequency process generated by Eq. \eqref{eq:2}.


\begin{table}[tbp]
    \centering
    \renewcommand{\arraystretch}{1.5} 
    \caption{Model parameters. Name, support and  interpretation.}
    \label{tab:pars}
    \begin{tabularx}{\linewidth}{|c|c|p{10.15cm}|}
        \toprule
        \textbf{Parameter} & \textbf{Support} & \textbf{Interpretation}  \\
        \midrule
        $A$ & $(0, \infty)$ & Amplitude for the sine wave. \\
        $B$ & $(0, \infty)$ & Amplitude for the cosine wave.  \\
        $b$ & $[0, 2 \pi)$ & Phase-offset of the signal. \\  
        $\sigma^2$ &  $(0, \infty)$ & Variance of measurements \\
        $a$ & $(0, \infty)$ & Mean of instantaneous growth rate \\   
        $\beta$ & $(0, \infty)$ & Adjustment rate of instantaneous growth rate  \\ 
        $\rho$ & $(0, 1)$ & One-lag autocorrelation of instantaneous growth rate, $\rho = \exp(-\Delta \beta)$  \\  
        $\omega^2$ & $(0, 2a\beta)$ & Infinitesimal variance of instantaneous growth rate  \\       
        $\gamma^2$ & $(0, a^2)$ & Stationary variance of instantaneous growth rate, $\gamma^2 = a\omega^2 / 2 \beta$ \\
        \bottomrule
    \end{tabularx}
\end{table}

\section{Estimation}\label{sec:estimation}

The estimation method has three steps, a preprocessing step, an initialization step (section \ref{sec:init}) and the main algorithm (section \ref{sec:saem}, \ref{sec:martingale} and \ref{sec:SMC}) We finally consider how to validate the model and quantify uncertainty using bootstrapping (section \ref{sec:bstrap}).

\subsection{Preprocessing} \label{ssec:preproc} \textit{Trend removal:} If the data show clear trends, these should be removed before the analysis. For our data, we only centered the signal around $y=0$ by subtracting the average $\bar y$. \textit{Amplitude normalization:} To obtain approximately constant amplitudes, we assume that the signal (roughly) concentrates around a narrow band of frequencies and that the amplitude process is slowly evolving, in order to use the Hilbert transform to compute the envelope of the signal \citep[p.~318--320]{hilbert}. We then apply a rolling average with a window size of approximately $10\%$ of the data to smooth the upper envelope, which we denote $z_i = z(x_i)$. The signal used for further analysis is then normalized as $y_i \leftarrow (y_i - \bar y)/z_i$. This implies that $\max (f(x)) =A+B \approx 1$, and we therefore assume that $B = 1-A$. After the analysis, the fitted curve can be transformed back to obtain a fit to the original data.



\subsection{Parameter initialization} \label{sec:init}

The loss function will likely exhibit many local maxima due to the trigonometric functions, and proper initialization will therefore assist the algorithm in convergence towards a global optimum. Obtaining good initial estimates is a non-trivial task, and typically involves heuristic approaches. Here we present our chosen initialization, denoted $\hat \theta^0$. In steps (1) and (2) we use a smoothed signal, denoted $\tilde y$, by performing loess regression with a span equal to 2\% of the data. This is implemented with the \texttt{loess} function in base R. 

\textit{(1) Amplitudes $A, B$ and phase offset $b$.} The signal \eqref{eq:1b} attains maximum and minimum values $f_{max} := A+B$ and $f_{min} :=-B-A^2/(8B)$ assuming, in case of the minimum, that $A\leq 4B$ (see supplementary material). Values $f_{max}$ and $f_{min}$ are unknown, but can be approximated by $\max(\tilde y) = \tilde y_{max}$ and $\min(\tilde y) = \tilde y_{min}$. Solving for $A$ and $B$, we obtain $\hat B = \frac{1}{9} \left( \tilde y_{max} - 4\tilde y_{min} \pm \sqrt{(\tilde y_{max} - 4 \tilde y_{min})^2 - 9 \tilde y_{max}^2} \right)$ and $\hat A = \tilde y_{max} - \hat B$, which offer two possible solutions, $(\hat A_1, \hat B_1)$ and $(\hat A_2, \hat B_2)$. Define the set of amplitude candidates 
\begin{equation} S = \{(A,B) \in ((\hat A_1, \hat B_1), (\hat A_2, \hat B_2), (\tilde y_{max}, 0))\, :\, A > B\}, \label{eq:ampcand} \end{equation}  
where we added $(\tilde y_{max}, 0)$ to ensure $S$ is non-empty. We initialize $\hat A^0, \hat B^0$ by drawing a random pair from $S$. The phase offset is initialized to $\hat b^0 = \pi$ to ensure that the initial grid of $b$ candidates span the entire unit circle in the first run of the algorithm, see Section \ref{sec:SMC}. 

\textit{(2) Measurement error: $\sigma^2$}. The variance of the measurement noise is approximated from the smoothed signal $\tilde y$, setting $(\hat \sigma^2)^0 = \frac{1}{n} \sum_{i=1}^n (y(x_i)-\tilde y(x_i))^2$. 

\textit{(3) Hidden process parameters: $a, \beta,\omega^2$}. 
With initial estimates for all other parameters, we now run SMC (Algorithm \ref{alg:SMC}) with a fixed number of particles $n_p$, but where each particle is initialized with a random pair $(a, \beta, \omega^2) \in (\tfrac{2\pi C_{min}}{n \Delta}, \tfrac{2\pi C_{max}}{n \Delta}) \times (0.01, 0.5) \times (0.01, 0.3)$ obtained from rejection sampling under the Feller constraint $2a\beta > \omega^2$. Here, $C_{min/max}$ is our prior belief on the lower/upper number of cycles in the signal. We resample among the particles at each step and select a final candidate based on the final weights. The parameters associated with the selected particle define our initial estimates $(\hat a^0, \hat \beta^0,(\hat \omega^2)^0)$. 

\subsection{Principle of the SAEM algorithm} \label{sec:saem}

Here, we propose an algorithm for estimating $\theta = (A, B, a, b, \sigma^2,\beta, \omega^2)$ from observations $y = (y_0, y_1, \ldots , y_n)$. 

The likelihood is not explicit as model \eqref{eq:1b} depends on the hidden stochastic process $(\xi_x)_{x \geq 0}$. 
A common approach for estimating in models with unobserved variables is the \textit{expectation-maximization} (EM) algorithm \citep{17}. In this iterative procedure at iteration $m$, the E step is carried out by evaluating the expected log-likelihood given data and current estimates $\hat{\theta}^{m-1}$: 
\begin{eqnarray}
Q_m(\theta) &\overset{def}{=}& Q(\theta \, | \, \hat \theta^{m-1}) \notag \\
&\overset{def}{=}& \E_{ \xi \, | \, y, \hat \theta^{m-1}}[\ell(y,\xi; \theta) \, | \, (y ; \hat \theta^{m-1}) ],\label{eq:EM1}
    \label{eq:Q}
    \end{eqnarray} 
where $\ell$ is the complete log-likelihood of $y$ and the hidden process $\xi = (\xi_0, \ldots, \xi_n)$. The expectation is taken with respect to the probability distribution $p(\xi | \, y, \hat \theta^{m-1})$. In the M-step, $Q_m(\theta)$ is then maximized in order to obtain new estimates $\hat \theta^{m}$: 
\begin{equation}
    \hat \theta^{m} =\argmax_{\theta} Q(\theta \, | \, \hat \theta^{m-1}).
    \label{eq:M}
\end{equation}
The complete log-likelihood is explicit and given by 
\begin{align}
\ell( y, \xi;\theta) &= \sum_{i=0}^{n} \log p(y_i | \xi_i) + \sum_{i=1}^{n} \log p(\xi_i \, | \, \xi_{i-1}) + \log p(\xi_0) \nonumber \\
&= -\frac{1}{2\sigma^2} \sum_{i=0}^{n} (y_i - f(x_i, \theta))^2 - \frac{n+1}{2}\log \sigma^2 
+ \sum_{i=1}^{n} \log p(\xi_i \, | \, \xi_{i-1}, \theta) + \log p(\xi_0). \label{eq:loglik}
\end{align}
The transition density $p(\xi_i \, | \, \xi_{i-1}, \theta)$ is  a non-central $\chi^2$-distribution, see Section \ref{sec:model}. 

The conditional distribution of $\xi$ given $y$ is not explicit due to the nonlinear nature of the regression function and the non-central $\chi^2$ distribution of the transition density. Thus, we cannot perform the E step in Eq. \eqref{eq:Q}. Instead, we use a variant of the \textit{Stochastic Approximation} EM (SAEM) algorithm \citep{53,16}, the novelty being the introduction of a martingale estimating function to solve the M step. In this procedure, the E step is replaced by two steps, a simulation (S) step and a stochastic approximation (SA) step. In the simulation step, we use a Sequential Monte Carlo (SMC) \citep{44, 56} sampler to draw the non-observed data $\xi^m$, conditionally on $\hat \theta^{m-1}$. We then apply a stochastic approximation to Eq. \eqref{eq:EM1}:
\begin{equation}
\tilde Q_m(\theta) = \tilde Q_{m-1}(\theta) + \alpha_m(\ell(y, \xi^m; \hat \theta^{m-1}) - \tilde Q_{m-1}(\theta)), \label{eq:EM2}
\end{equation}
where $\alpha_m$ is a decreasing sequence of positive numbers quantifying the memory in the approximation process, fulfilling $\sum_m \alpha_m = \infty$ and $\sum_m \alpha_m^2 < \infty$, which ensures convergence of the algorithm \citep{53}.

The algorithm can be simplified when the complete likelihood belongs to a curved exponential family, which then reduces to approximating the minimal sufficient statistics of the model during the E step, while the M step is explicit through the sufficient statistics. Nevertheless, the distribution $p(\xi_i \, | \, \xi_{i-1})$ does not belong to a curved exponential family, and we do not have sufficient statistics. In that case, each iteration of the EM algorithm solves
$$\partial_\theta \tilde Q_m(\theta) = 0.$$
However, solving the score function for the square-root process can be numerically unstable, because it involves a Bessel function of the first kind \citep{DitlevsenLansky2006}. The square-root diffusion belongs to the class of Pearson diffusions, which are statistically tractable and well-behaved \citep{FormanSorensen2008}.  Martingale estimating functions provide a useful alternative with well established asymptotic properties (consistency, asymptotic normality) similar to the maximum likelihood estimator \citep{soerensen2012}. Furthermore, estimating functions have shown useful for parameter estimation in the square-root process when only observed through the integrated process \citep{38}, which is the case of the growth process $g$. Therefore, we propose to use a martingale estimating function \citep{45, soerensen2012} to solve the M step, leading to an approximate version of the EM algorithm.

\subsection{Martingale estimating function and SAEM-variant} \label{sec:martingale}

At iteration $m$ of the algorithm, given the current estimate $\xi^{m-1}$ of the hidden process $\xi$ obtained using SMC (Section \ref{sec:SMC}), the M step is updated by solving the score function, i.e., by differentiating
the complete log-likelihood $\ell(y, \xi^{m
}, \theta)$ with respect to $\theta$, the score function being 
approximated during the SA step. 

The estimators of parameters $a, \beta$ and $\omega$ of the hidden process $\xi$ are defined directly as the solution to the martingale estimation function, mimicking the score equation. This leads to the statistics $S_1,S_2,S_3$  \citep[p. 21]{soerensen2012}:
\begin{align}
&S_1 = \dfrac{\frac{1}{n} \sum_{i=1}^n \xi_{i \Delta}^{m}/\xi_{(i-1) \Delta}^{m}-(\frac{1}{n} \sum_{i=1}^n \xi_{i \Delta}^{m})(\frac{1}{n} \sum_{i=1}^n(\xi_{(i-1) \Delta}^{m})^{-1})}{1- (\frac{1}{n} \sum_{i=1}^n \xi_{(i-1)\Delta}^{m})(\frac{1}{n} \sum_{i=1}^n(\xi_{(i-1)\Delta}^{m})^{-1})};\\
&S_2 = \frac{1}{n} \sum_{i=1}^n \xi_{i \Delta}^{m} + \dfrac{S_1}{n(1-S_1)} (\xi_{n\Delta}^{m} -\xi_0^{m});\\
&S_3 = \dfrac{\sum_{i=1}^n (\xi_{(i-1) \Delta}^{m})^{-1}(\xi_{i \Delta}^{m} - \xi_{(i-1) \Delta}^{m}S_1 - S_2 (1-S_1))^2}{\sum_{i=1}^n (\xi_{(i-1)\Delta}^{m})^{-1}((\frac12 S_2 - \xi_{(i-1)\Delta}^{m})S_1^2-(S_2 - \xi_{(i-1)\Delta}^{m})S_1+\frac12 S_2)}.
\end{align}

The parameter  $\sigma^2$ is easier to obtain, as the derivative of the complete log-likelihood leads to the following statistic:
\begin{align}
    & S_4 = \frac{1}{n}\sum_{i=0}^n (y_i - f(x_i, \hat\theta^{m-1} \, | \, \xi^{m}))^2,
\end{align}
where $g^m(x_i) = \int_0^{x_i} \xi^{m}_s d s$.
These statistics are then approximated by the stochastic approximation scheme, as in the standard SAEM. Following \cite{30,40}, we choose the form 
\begin{equation}\alpha_m = \begin{cases} \label{eq:mem}
    1 & m \leq m_0 \\
    (m-m_0)^{-0.8} & m > m_0
\end{cases},
\end{equation} 
for the Stochastic Approximation sequence in Eq. \eqref{eq:EM2}, where $m_0$ determines the first iteration that includes memory from the previous step. We obtain approximations of the statistics,  
\begin{equation}
    s_k^{(m)}   = s_k^{(m-1)} + \alpha_m (S_k -s_k^{(m-1)}), \quad k=1, \ldots, 4.
\end{equation}
We then update the parameter estimates in terms of these statistics
\begin{align*}
& \hat \rho^{m} = s_1^{(m)}; \quad \hat \beta^{m} = -\frac{1}{\Delta} \log(\hat \rho^{m}); \quad \left ( \hat \omega^{m}\right )^2  = s_3^{(m)}; \quad \left ( \hat \sigma^{m} \right )^2 = s_4^{(m)};  \\
&\hat a^{m} = s_2^{(m)}\mathbbm{1}_{\{ 2 \hat a^0 > s_2^{(m)} > \frac 12 a^0 \}} + \hat a^{m-1} \left(1-\mathbbm{1}_{\{ 2 \hat a^0 > s_2^{(m)} > \frac 12 a^0 \}} \right).
\end{align*}

The number of cycles is proportional to $a$. We ensure that the process does not mistake random fluctuations for cycles by restricting estimates of $a$ to be between half and double of the initial estimate (see Section \ref{sec:init}). If this condition is satisfied for the statistic associated with $a$, we update, otherwise the estimate from the previous step is kept. This can, of course, be adapted to the requirements of specific applications.

To update $A$ and $B$, we use that after normalization $B = 1-A$. Substituting in Eq. \eqref{eq:1b} we obtain
\begin{equation} f(x_i, \theta) = Aw_i + v_i, \label{eq:Areg}\end{equation}
where $w_i = \sin(g(x_i)+b) + \cos(2g(x_i)+2b)$ and $v_i = - \cos(2g(x_i)+2b)$. Eq. \eqref{eq:Areg} allows for a linear regression with constraint $A \in (0.5, 1)$ since $A+B=1$ and $A>B > 0$. Then

\begin{align}
    \hat A^{m} 
    &= \argmin_{A \in \left( 0.5, 1 \right)} \sum_{i=1}^n \left(y_i - A w_i - v_i\right)^2 \\
    &= \min\left(1,\; \max\left( 0.5,\; \frac{\sum_{i=1}^n w_i (y_i - v_i)}{\sum_{i=1}^n w_i^2} \right) \right), \label{eq:Ahat}
\end{align}
where $\sum_{i=1}^n w_i (y_i - v_i)/\sum_{i=1}^n w_i^2$ is the least squares estimator in the unconstrained problem. Finally, $\hat B^{m} = 1 -  \hat A^{m}$. 

It turns out that the initial phase $b$ and the growth process $g(x)$, and consequently $\xi$, are not individually identifiable and must be estimated jointly. In section \ref{sec:SMC}, we go into detail on how we can modify the SMC sampler to simultaneously sample $\hat b^{m}$ in iteration $m$, by making a grid of plausible $b$ candidates, and run a single SMC filter for each $b$. We call this SMC$^+$. A brief example showing why sequential estimation of $g(x)$ and $b$ is not feasible is given in the supplementary material.

The SAEM algorithm iterates until some stopping criteria is met. Several rules exist for determining when convergence can be assumed \citep{21, 22, 52, 53}. One common rule is to ensure that the maximum of the relative differences of all parameters between consecutive iterations falls below some threshold to ensure that the SAEM estimates have stabilized. We took the threshold to be $10^{-4}$ following \cite{54}. We also make a simulation study to investigate the sensitivity to this threshold choice, see supplementary material.

In summary, the algorithm is as follows. We first use SMC to filter multiple trajectories of $\xi$ called particles, using the transition density. This is repeated over a grid of candidates for the phase parameter $b$. This algorithm is detailed in section \ref{sec:SMC}. Once the latent process is filtered, we proceed to the stochastic approximation step. 
In the final maximization (M) step, we update $\hat \theta^{m}$. This procedure is repeated until some stopping criteria are met. 
The basic estimation procedure is summarized in Algorithm \ref{alg:cap}. Later we construct confidence intervals for the parameters and other derived quantities. This is summarized in Algorithm \ref{alg:bs}.

\begin{algorithm}[H]
\linespread{1.35}\selectfont
\SetAlgoLined
\caption{Parameter estimation using SAEM and martingale estimating functions}\label{alg:cap}
\KwData{$(X,Y) := \{(x_1,  y_1), \hdots, (x_n,  y_n)\}$} 
\KwResult{${\xi}, \hat{\theta} = ( \hat{A}, \hat B,  \hat{b}, \hat{a}, \hat{\sigma}^2, \hat \rho, \hat{\omega}^2)$}
\texttt{(1)} \,  \,Initialize parameters $\hat{\theta}^0$ and the unknown process ${\xi}^0$ (Section \ref{sec:init})\;
$m \gets 1$ \;
\While{\textrm{"Stopping criteria"}}{
\texttt{(2)} \,  Update $\hat b^m, \xi^m$ given $\hat{\theta}^{m-1}$ using SMC$^+$ (Algorithm \ref{alg:SMC}) \;
\texttt{(3)} \, Do the stochastic approximation of   statistics $S_1, \ldots, S_4$\;
\texttt{(4)}  \, Update $\hat a^m, \hat \rho^m, (\hat \omega^2)^m,  (\hat \sigma^2)^m$ using the  statistics $s_1^{(m)} \ldots, s_4^{(m)}$\;
\texttt{(5)}  \, Update $\hat A^m$ using Eq. \eqref{eq:Ahat} and set $\hat B^m = 1 - \hat A^m$ \;
$m \gets m+1$ 
}
\end{algorithm}

\subsection{SMC algorithm}\label{sec:SMC}

In Algorithm \ref{alg:cap} we apply SMC to filter out the hidden process $\xi$. 
The SMC algorithm provides a set of $n_p$ particles $(\xi^{(j)})_{j=1, \ldots, n_p}$ and weights $(W^{(j)})_{j=1, \ldots, n_p}$ approximating the conditional smoothing distribution $p(\xi_0, \ldots, \xi_n|y_0, \ldots, y_n; \theta)$ \citep{41,42}. The SMC algorithm is an iterative algorithm. At each iteration, particles are sampled from a proposal distribution  $q(\xi_i|\xi_{i-1}, y_i, y_{i-1})$.  To ease notation, we denote $\xi_{0:i}=(\xi_0, \ldots, \xi_i)$ and likewise for $y_{0:i}$. 

We choose the transition density $p(\xi_i|\xi_{i-1})$ (see Section \ref{sec:model}) as proposal $q$. This choice simplifies the weight to
\begin{equation}
    W = \frac{p(y_i, \xi_{0:i} \, | \, y_{0:i-1},  \theta)}{q(\xi_i \, | \, y_i, \xi_{0:i-1},  \theta)} = p(y_i \, | \,  \xi_{i},  \theta) \sim \mathcal{N}(f(x_i, \theta), \sigma^2).
\end{equation}
The basic SMC algorithm is presented as pseudocode in Algorithm \ref{alg:SMC} for a given value $\theta$. It provides an empirical measure which is an approximation of the smoothing distribution $p(\xi_{0:n}|y_{0:n}, \theta)$. A sample from this empirical distribution is obtained by sampling an index $j$ from a multinomial distribution with probabilities $(W^{1}_n, \ldots, W^{n_p}_{n})$. 

Phase offset $b$ and the growth process $g(x)$ are intrinsically connected and need to be handled jointly. We therefore propose to use SMC with a grid search adaptation, denoted SMC$^+$, described in Algorithm \ref{alg:SMCplus}. SMC$^+$ uses equally spaced candidates of $b$, centered around the current estimate of $b$, and runs the basic SMC with each candidate. We consider $G = 20$ equidistant values. We set a range $2w$ around the current estimate to search. Initially, we set $w = \pi$, which decreases as the Stochastic Approximation sequence in Eq. \eqref{eq:mem} decreases, resulting in a grid that collapses around the current estimate of $b$. 

\begin{algorithm}[htb]
\linespread{1.35}\selectfont
\SetAlgoLined
\caption{Sequential Monte Carlo (SMC)}\label{alg:SMC} 
\KwData{$(X,Y)$, $\theta = ( {A}, {B},  {a}, {b}, {\sigma}^2, {\beta},  \rho, {\omega}^2)$} 
\KwResult{$ (\xi^{(j)} = (\xi^{(j)}_0 , \ldots, \xi^{(j)}_n))_{j=1, \ldots, n_p}$}
\texttt{(1)} \,Initialize $n_p \times n+1$ matrices $\xi$ and weights $W$, where $n_p$ is the number of particles\;
\texttt{(2)} \, For $j=1, \ldots, n_p$: set $\xi^{(j)}_0 = a$ and $W^{(j)}_0 = 1/n_p$\;

\For{$i = 1$ \KwTo $n$}{
    \texttt{(3)} \,Set $\lambda_i = 2  \rho c \, \xi^{(j)}_{i-1}$ for  $j=1, \ldots n_p$\; 
    \texttt{(4)} \,Update particles: $\xi^{(j)}_i \sim \frac{1}{2c} \chi'^2_{\nu}(\lambda_i) $ for $j=1, \ldots n_p$\;
    \texttt{(5)} \,Compute weights: $W^{(j)}_i = p(y_j \, |  \xi^{(j)}_i , \theta)$ for $j=1, \ldots n_p$\;
    \texttt{(6)} \,Normalize weights: $W^{(j)}_i = W^{(j)}_i/ \sum_{i=1}^{n_p} W_i^{(j)}$ for $j=1, \ldots n_p$\;
    \texttt{(7)} \,Resample particles by drawing $n_p$ indices from the set $\{1, \ldots, n_p\}$ with probabilities $W_i^{(1)}, \ldots, W_i^{(n_p)}$. Denote the realizations $\{I_1, \ldots, I_{n_p}\}$ and set $\xi_{0:i}^{(j)} = \xi_{0:i}^{(I_j)}$\;
    
}
\end{algorithm}

\begin{algorithm}[htb]
\linespread{1.2}\selectfont
\SetAlgoLined
\caption{SMC$^+$: Grid Search}\label{alg:SMCplus}
\KwData{$(X,Y)$, $\theta = ( {A}, {B}, b, {a}, {\sigma}^2, {\beta},  \rho, {\omega}^2)$}
\KwResult{$\hat b,  \hat \xi$}
 \texttt{(0)} \, Set number of candidates $G$ (grid granularity);
\texttt{(1)} \, Define half‐width
\[
w \;=\;
\begin{cases}
\pi, & m \leq m_0,\\
\pi \times (m - m_0)^{-0.8} & m > m_0.
\end{cases}
\]
\texttt{(2)} \, Form grid of $G$ phase zero candidates:
\[
b_j = \bigl(b - w + (j-1)\tfrac{2w}{G-1}\bigr)\bmod 2\pi,\quad j=1,\dots,G.
\]
\For{$j = 1$ \KwTo $G$}{
 \texttt{(3)} Run SMC (Algorithm \ref{alg:SMC}) with $\theta$ and $b_j$ replacing $b$, to obtain $\hat \xi^{(j)}$\;
 \texttt{(4)} Given $\theta$, with $b_j$ replacing $b$, and $\hat \xi^{(j)}$ compute the log likelihood \eqref{eq:loglik} and denote it $L_j$\;
}
\texttt{(5)} \, Set $(\hat b, \hat \xi) = (b_{j^*}, \xi^{(j^*)})$ where $L_{j^*} \geq L_{j}$ for $j = 1,\ldots, G$.

\end{algorithm}



\subsection{Bootstrapping} \label{sec:bstrap}


Standard errors and confidence intervals for all parameters estimated using SAEM (Algorithm \ref{alg:cap}) are computed using residual bootstrapping. The general procedure to obtain a set of bootstrapped estimates is described in Algorithm \ref{alg:bs}. From these relevant summary measures can be computed.

\begin{algorithm}[H]
\linespread{1.35}\selectfont
\SetAlgoLined
\caption{Residual Bootstrapping (RB)}\label{alg:bs}
\KwData{Observations $y = (y_1, \ldots,y_n)$ and fitted values $\hat y = (\hat y_1, \ldots ,\hat y_n)$} 
\KwResult{Vector of estimates $( \hat \theta_m )_{m=1, \ldots,M}$}
$m \gets 1$ \; 
\texttt{(1)} \,Set number of bootstrap replications $M$\; 
\texttt{(2)} \, Compute residuals $r_i = y_i - \hat{y_i}$, for $i=1, \ldots, n$\;
\While{m < M}{
\texttt{(3)} \, Let $r_i^{(m)}$ be a random sample from the set of residuals $\{r_1, r_2, \ldots, r_n\}$, for $i=1, \ldots, n$\;
\texttt{(4)} \, Construct replicate signal as $y_i^{(m)} =\hat{y}_i+r_i^{(m)}$, for $i=1, \ldots, n$\;
\texttt{(5)} \, Estimate $\hat \theta_m$ using Algorithm \ref{alg:cap} on the replicate $(y_1^{(m)}, \ldots, y_n^{(m)})$ \;
$m \gets m+1$ 
}
\end{algorithm}

\subsection{Model validation}\label{sec:residuals}

To assess the quality of the model, we use the raw residuals, 
\begin{equation} \label{eq:resdef}
    \textrm{res}_i = y_i - \hat y_i,
\end{equation}
where $\hat y_i$ is the fitted value of observation $i$. We validate the mean structure and homoscedasticity assumption by residual plots. Additionally, we compare the empirical residuals \eqref{eq:resdef} with theoretical normally distributed errors. 

To address the estimator variability and bias, we compute the (vector of) relative differences 
between the estimated and bootstrapped versions: $(\tilde \theta -  \hat \theta_m)/\hat \theta$. 

\section{Simulation study}\label{sec:simulation}

In this Section, we simulate a collection of diverse signals - varying in parameter configurations and lengths - and evaluate both the accuracy of the estimates and the precision with which we recover the number of cycles  \eqref{eq:numberofcycles}.

A total of 800 signals were simulated, with a fixed step size equal to $\Delta = x_i - x_{i-1} = 1$ and individual parameters $\theta_m, m = 1, \ldots , 800$. 
To obtain the hidden stochastic process, the solution to Eq. \eqref{eq:2}, we simulated trajectories using the Euler-Maruyama scheme with step size $\Delta/100$ starting at $\xi_0 = 0$. We then obtained $g(x_i) = g(x_{i-1}) + \sum_{j=1}^{100} \xi_{(i-1)j} \Delta/100$, where $\xi_{ij}$ is the simulated value at $x=i + j/100$. Finally, we sampled a signal using Eqs. \eqref{eq:1} and \eqref{eq:1b}.

The simulation details are as follows. We simulated 200 data sets for each of 4 sample sizes; short ($n=200$), medium short ($n=400$), medium long ($n=600$) and long ($n = 800$). We set $C_{\min} = 2$ and $C_{\max} = \max(6, n/100 \times 5)$ to be the minimum and maximum of allowed number of cycles. 
The upper and lower frequency is then set to $a_{\min}  = 2 \pi C_{\min}/N \Delta$ and $a_{\max} = 2 \pi C_{\max}/N \Delta$. The values of the parameters were sampled from uniform distributions $\mathcal{U}$ with the following supports,

\begin{align} \label{eq:10}
    &\beta \sim \mathcal{U}(0.01, 3) , \quad a \sim \mathcal{U}(a_{min}, a_{max}) , \quad \omega \sim \mathcal{U}(0.01, 0.3) \\[1ex]
    &\sigma \sim \mathcal{U}(0.2,0.6) , \quad b \sim \mathcal{U}(0, 2\pi), \quad  A \sim \mathcal{U}(0.5, 1),
\end{align}
and from this, $B = 1 - A$ and  $\rho = \exp(-\Delta \beta)$. The supports of the uniform distributions were chosen to best reproduce the real signals of the case study. For the process parameters in Eq. \eqref{eq:10} we did rejection-sampling, and only accepted the triple $(a, \beta, \omega)$ if they fulfilled the Feller condition $2a\beta > \omega^2$ to ensure that $\xi_x > 0$ for all $x$.

\begin{figure}[!tb]
    \centering
    \begin{minipage}{0.5\linewidth}
        \centering
        \includegraphics[width=\linewidth]{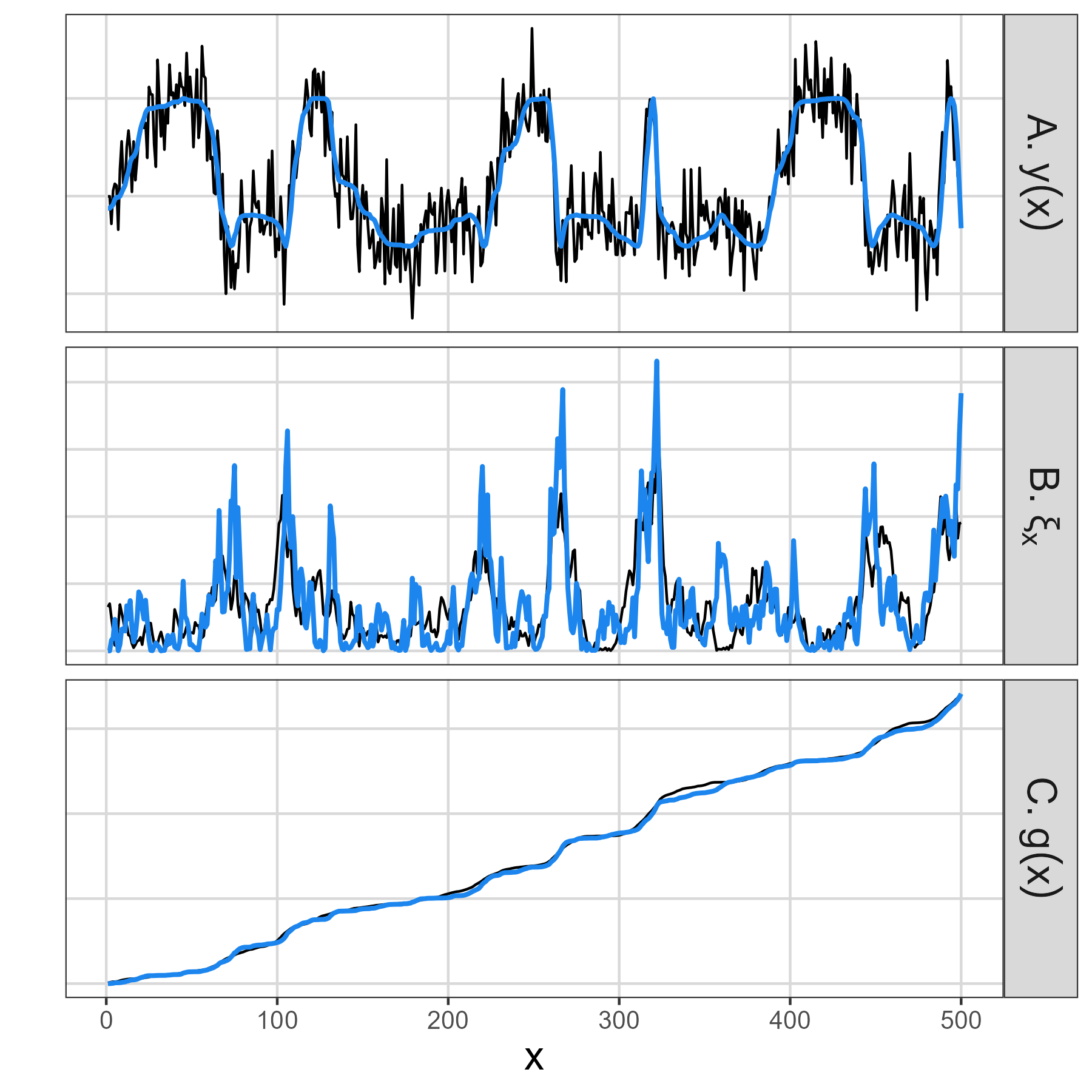}
    \end{minipage}%
    \hfill
    \begin{minipage}{0.45\linewidth}
        \caption{\textbf{Simulated and fitted signal}. Simulated signals are in black, fitted signals in blue. The dataset was simulated using $n=500$, $\Delta = 1$, $\beta = 0.07$, $\sigma = 0.3$,  $A = 0.6$, $B = 0.4$, $a=0.05$, $b = \pi/20$, $\omega^2 = 0.064$. The algorithm used $n_p = 1500$ number of particles. A. Simulated signal and fitted signal. B. Simulated hidden process $(\xi_x)_{x \geq 0}$ and estimate $(\hat \xi_x)_{x \geq 0}$. C. Simulated growth process $g(x)$ and estimate $\hat g(x)$.} 
        \label{fig:simfits}
    \end{minipage}
\end{figure}

A simulated signal is presented in the top plot of Figure \ref{fig:simfits}, which also displays the model fit. The second row shows the hidden process $(\xi_x)_{x \geq 0}$ underlying the simulated signal, both the true process and the estimated version. The third row displays the growth process $\hat g(x)$, along with its estimated counterpart.

\begin{figure}[!tb]
    \centering
    \includegraphics[scale=0.27]{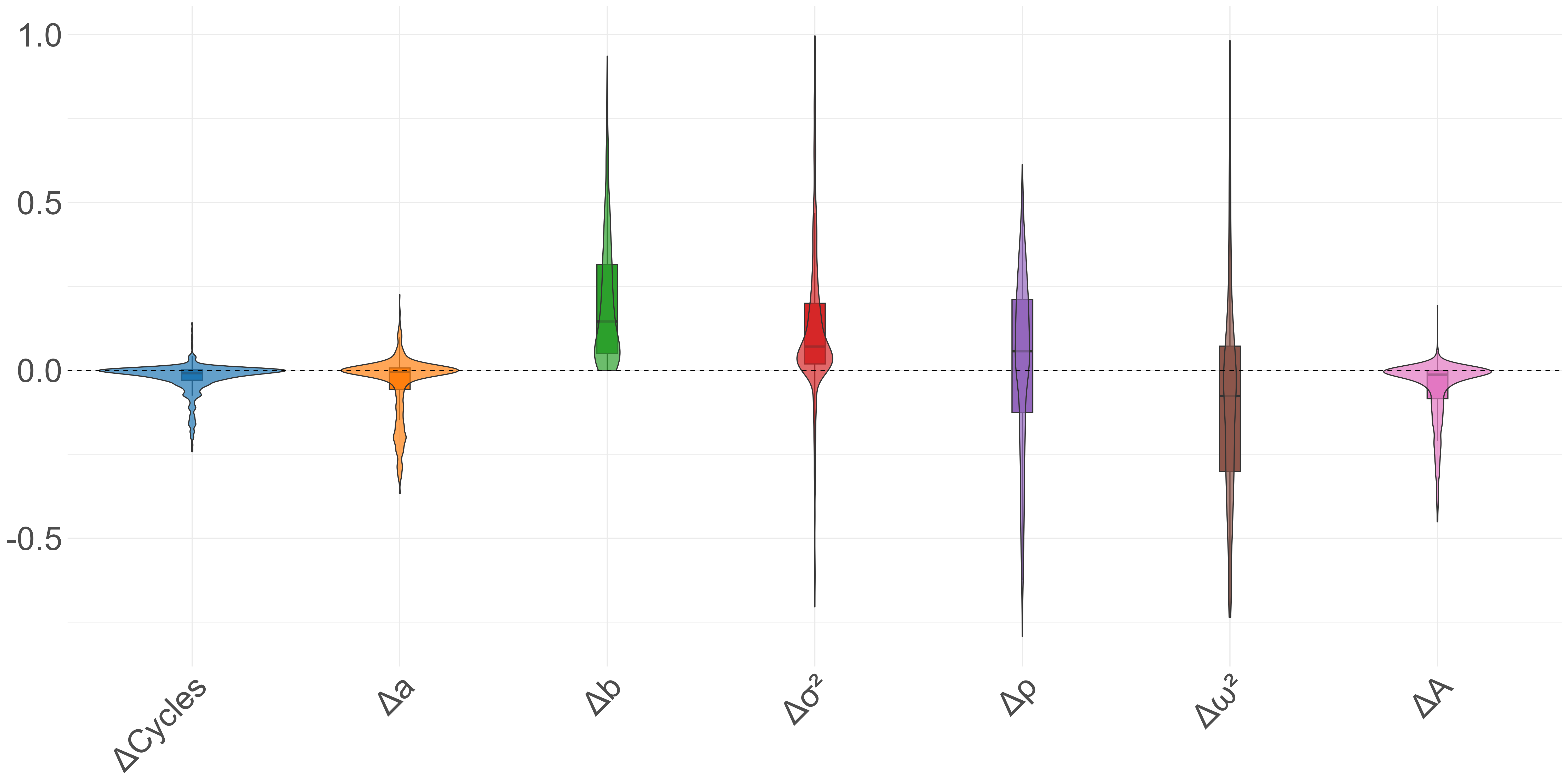}
    \caption{\textbf{Estimator variability.} $M=800$ simulations equally divided into 4 groups of different length ($n=200$, $n=400$, $n=600$, $n=800$) and with random parameter configurations (see Section \ref{sec:simulation}). Violinplots show the relative difference $(\theta_m -  \hat \theta_m)/ \theta_m, \, m = 1, \ldots , 800$ of parameter estimates $\hat \theta_m$ obtained from signals simulated with parameter $\theta_m$. The number of cycles are computed with Eq.  \eqref{eq:numberofcycles}. For $b$ we computed the smallest absolute relative difference on the unit circle. This was divided by $\pi$ to obtain a relative difference.}
    \label{fig:simviolin}
\end{figure}

\begin{figure}[!tb]
    \centering
    \begin{minipage}{0.75\linewidth}
        \centering
        \includegraphics[width=\linewidth]{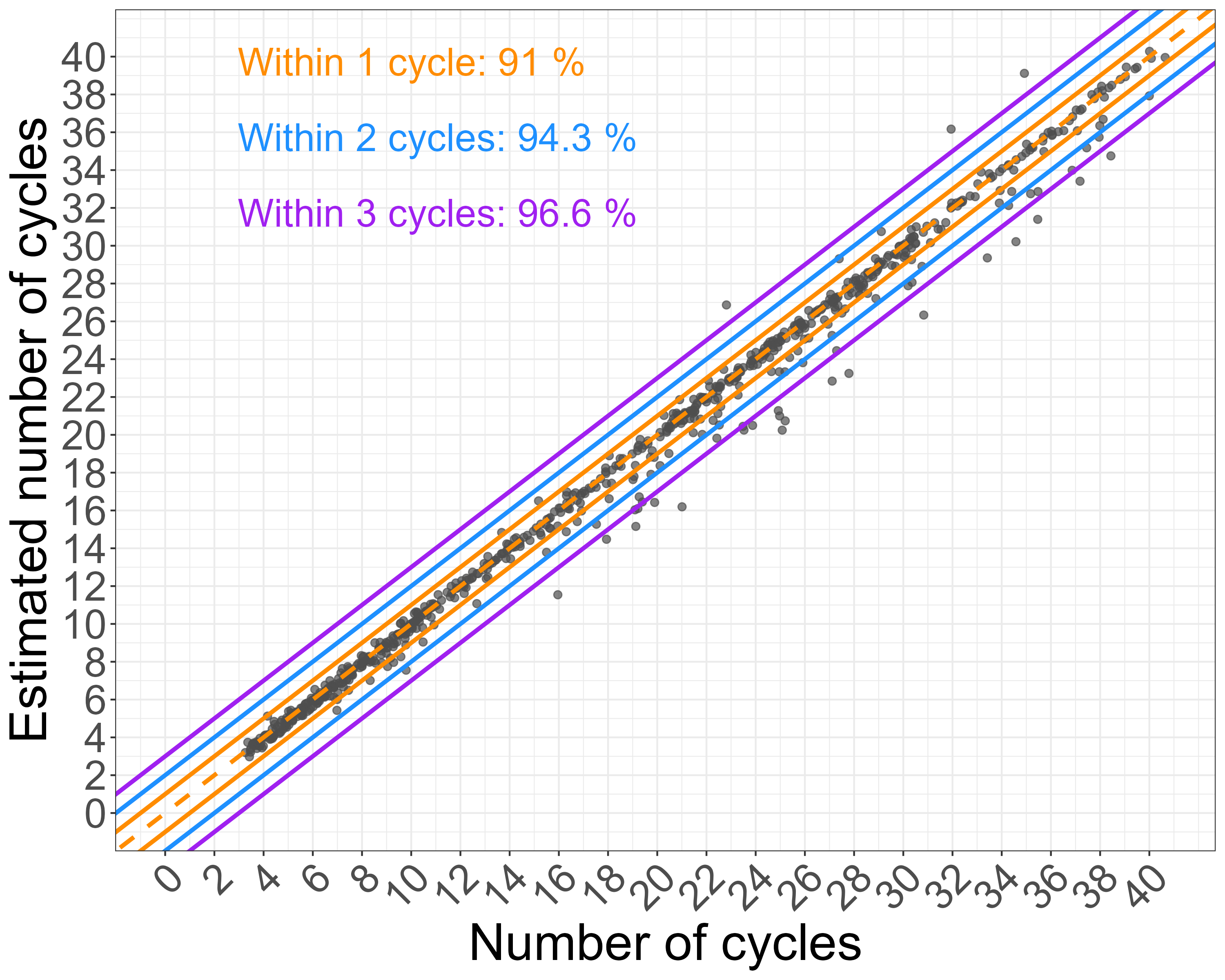}
    \end{minipage}%
    \hfill
    \begin{minipage}{0.2\linewidth}
        \caption{\textbf{Estimated number of cycles against true number of cycles}. The colored ribbons enclose estimates of number of cycles from simulated data sets with less than 1, 2, and 3 cycles of deviation from the true number of cycles.}
        \label{fig:agevsest}
    \end{minipage}
\end{figure}

For each simulated signal $m$, we computed the relative difference between the true parameter $\theta_m$ and its estimate $\hat \theta_m$ defined as $(\hat \theta_m - \theta_m)/ \theta_m$, including the number of cycles. Figure \ref{fig:simviolin} shows that the estimates of $a$ and $A$ are close to the true values relative to their scale, whereas the other parameters have larger variance, but without a prominent bias, maybe except for $b$. 
Most importantly, the number of cycles, Eq. \eqref{eq:numberofcycles},  displays low variance and low bias, which is one of the key questions, since it provides an age estimate. Figure \ref{fig:agevsest} shows that $91 \%$ of the estimates fall within 1 cycle from the true number of cycles, and less than $4 \%$ deviate by more than 3 cycles, which is mainly happening for signals with many cycles, where the peaks are more tightly concentrated.

In Section \ref{sec:case} we fit the model to data obtained from  measurements of isotopes within narwhal tusks.

\section{Age determination of a narwhal tusk} \label{sec:case}

 First we introduce the case study. The data is presented in Section \ref{ss:casedata}. In Section \ref{ss:dating} we discuss how the model and the data can be used to date the narwhals. In Section \ref{ss:context} we  provide the estimated age, along with a confidence interval and compare to estimates from manual counts. Section \ref{ss:modelval} presents model validation, checking if the data conform to the model assumptions.

\begin{figure}[!tb]
    \centering
    \begin{subfigure}[b]{0.45\linewidth}
        \includegraphics[width=\linewidth]{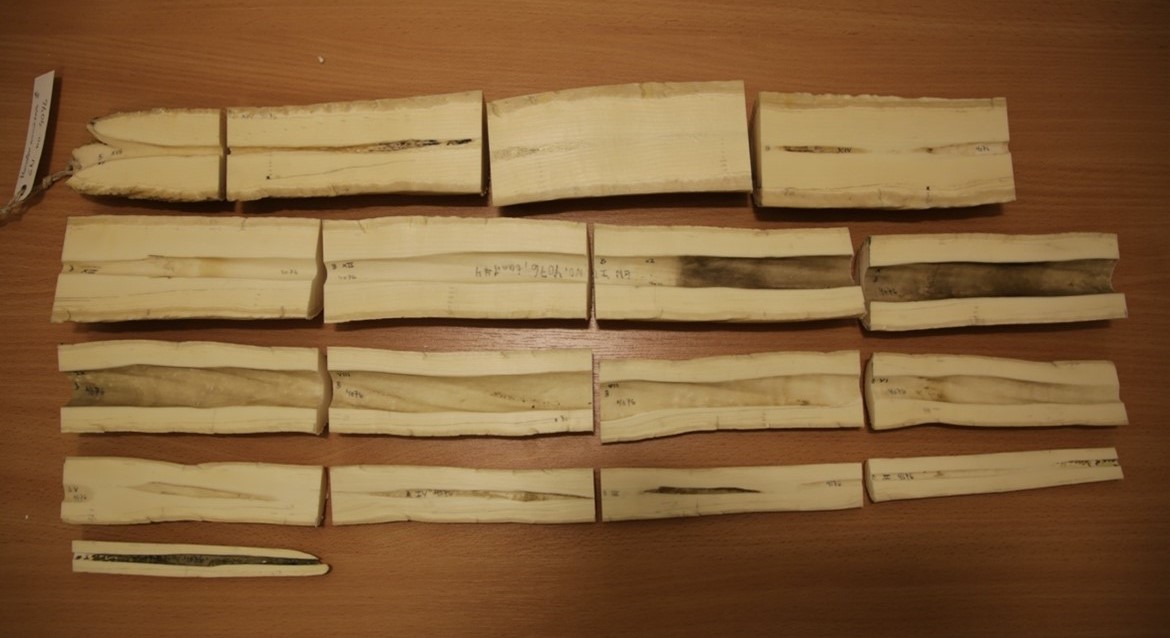}
        \caption{Narwhal tusk split in half, and cut into several pieces revealing the outer layer Cementum, the inner layer Dentin and the cavity Pulp.}
        \label{fig1:tusk}
    \end{subfigure}
    \hfill
    \begin{subfigure}[b]{0.45\linewidth}
        \includegraphics[width=\linewidth]{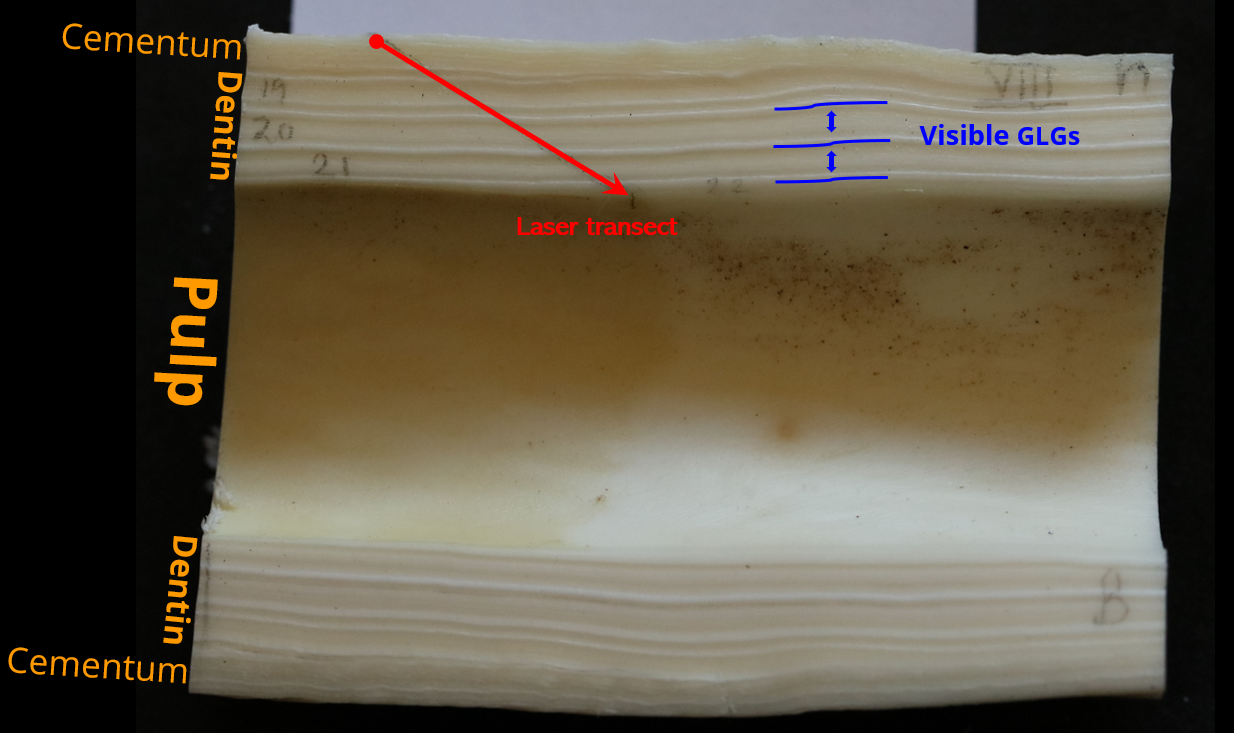}
        \caption{Focus on one single tusk piece, with clear Growth Layer Groups (blue lines), and highlighted transect which is used in the LA-ICP-MS analysis.}
        \label{fig2:tusk}
    \end{subfigure}
    \bigskip
    \begin{subfigure}[b]{\linewidth}
        \centering
        \includegraphics[width=0.8\linewidth]{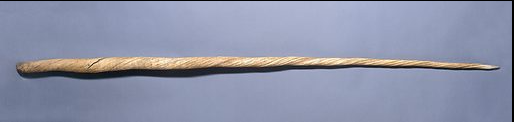}
        \caption{A tusk of a narwhal, with its signature left twisting appearance.}
        \label{fig:wholetusk}
    \end{subfigure}
    \caption{{\bf Tusk and tusk pieces}.}
\end{figure}

Narwhals exhibit a consistent and well-documented annual migration pattern as they move between specific summer and winter habitats, which are situated in fjords and inlets, and offshore areas, respectively \citep{1}. The primary prey items consist of halibut, squids, and polar cod \citep{3,4}. On their seasonal migration routes, the narwhals travel through the same regional water masses to and from summer and winter grounds. The strict diet and rigid migration pattern result in annual depositions of growth layers in the dentin zone of the tusk, consisting of an opaque and a translucent layer,  see Figure \ref{fig2:tusk}.
Continuous fluctuations in the deposition/biomineralization rates show up in the signals and likely occur due to seasonal variations driven by environmental factors, such as temperature or salinity \citep{25}. In addition, the physiological characteristics of the animals, notably age and health, likely modulate metabolic processes and as a result potentially influence the biomineralization rate \citep{26,27}. These assumptions are based on otoliths studies, and it seems reasonable to assume that the tusk growth rate exhibit similar characteristics.
The recurring deposition of GLGs in conjunction with a variable deposition rate produces a cyclic profile that fits well into our working model \eqref{eq:1}. Since the hypothesis is that one GLG is deposited annually, counting these cycles will theoretically provide an age estimate of the narwhal. An approximate determination of the range of GLGs in a tusk piece can be made through visual observation, see Figure \ref{fig2:tusk}. However, in practice visually reading of the GLGs is both difficult and highly uncertain, and automatic robust methods based on more objective criteria and quantitative measures are needed. 

\subsection{Data acquisition and preparation} \label{ss:casedata}

\begin{figure}[!tb]
    \centering
    \begin{minipage}{0.67\linewidth}
        \centering
        \includegraphics[width=\linewidth]{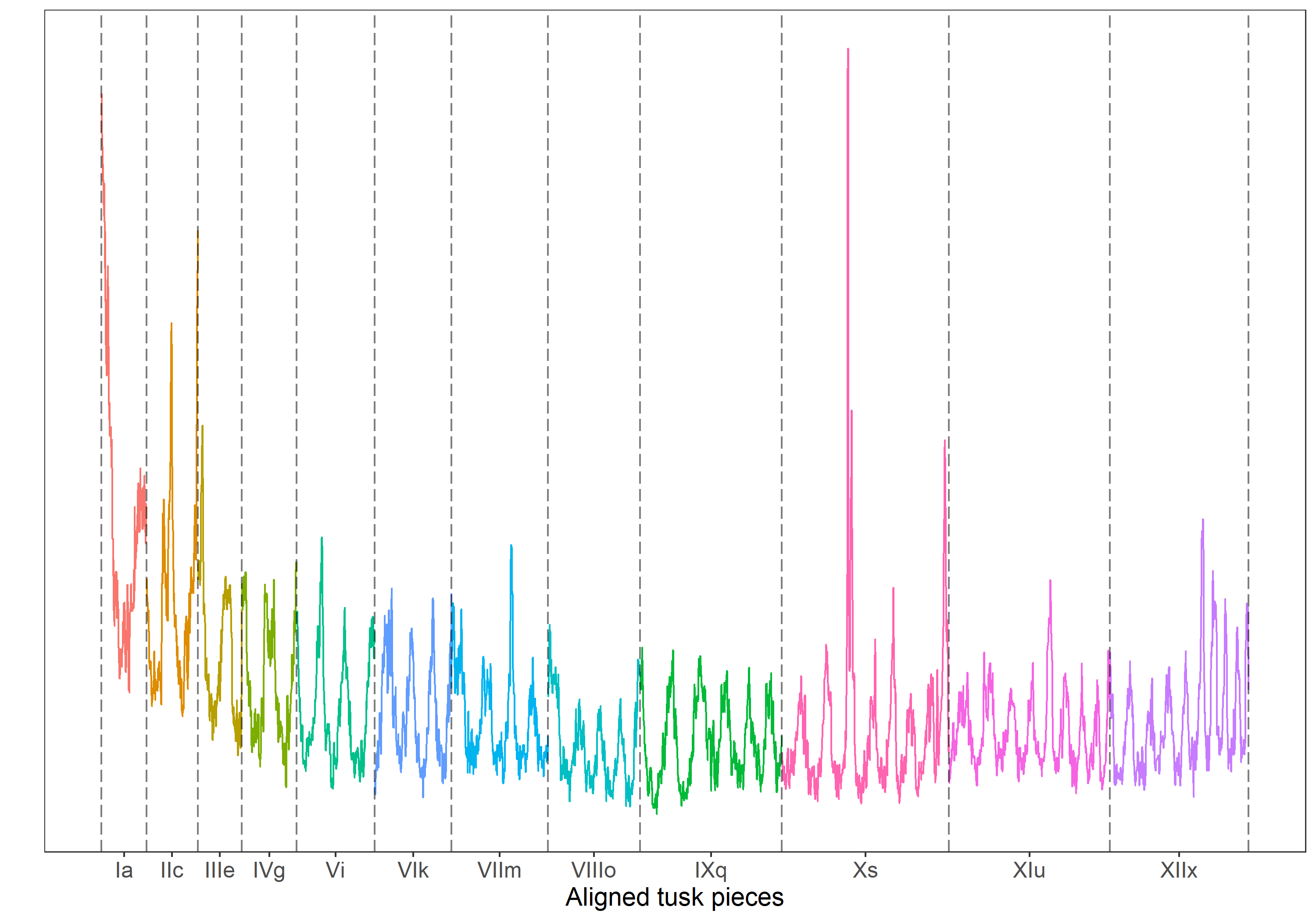}
    \end{minipage}%
    \hfill
    \begin{minipage}{0.28\linewidth}
        \caption{\textbf{Analyzed data set}. Calcium-43 adjusted Barium-137 signals obtained from tusk 956. The data set consists of 12 signals merged together into one common time series in chronological order. The first part of the time series corresponds to the tip of the tusk, deposited when the whale was born, and the last part of the series corresponds to the base nearest the skull, deposited just before death.}
        \label{fig:alltusks}
    \end{minipage}
\end{figure}

We analyzed one tusk (ID 956) collected from the Inuit Hunt of narwhals in Niaqornat, West Greenland, in 2010. The tusk was sectioned in two halves and one was used for subsequent analysis. The half tusk was divided into several tusk pieces covering all GLGs (see Figure \ref{fig1:tusk}). Few consecutive pieces had a complete overlap of GLGs, thus only one of these were used. After the exclusion, $12$ pieces remained. Each tusk piece was measured along a transect (see Figure \ref{fig2:tusk}) using Laser Ablation Inductively Coupled Plasma Mass Spectrometry (LA-ICP-MS) by the Geological Survey of Denmark and Greenland (GEUS), providing the concentrations of 14 isotopes and trace elements (Sr, Ba, Zn, Mg, K, Li, Mn, Pb, Cu, P, Ti, Cr, Co, Rb) \citep{28,29}. Our analysis centers on Barium-137 normalized by the isotope Calcium-43 in high abundance, since among the set of elements, Barium displayed distinct cyclic patterns speculated to correlate with migration between summer and winter grounds \citep{5,6}. Consequently two peaks in the signal would correspond to one annual cycle. The combined signal over all pieces is visualized in Figure \ref{fig:alltusks}.

\subsection{Dating the observations} \label{ss:dating}

\begin{figure}[!tb]
    \centering
    \begin{minipage}{0.60\linewidth}
        \centering
        \includegraphics[width=\linewidth,height=8cm]{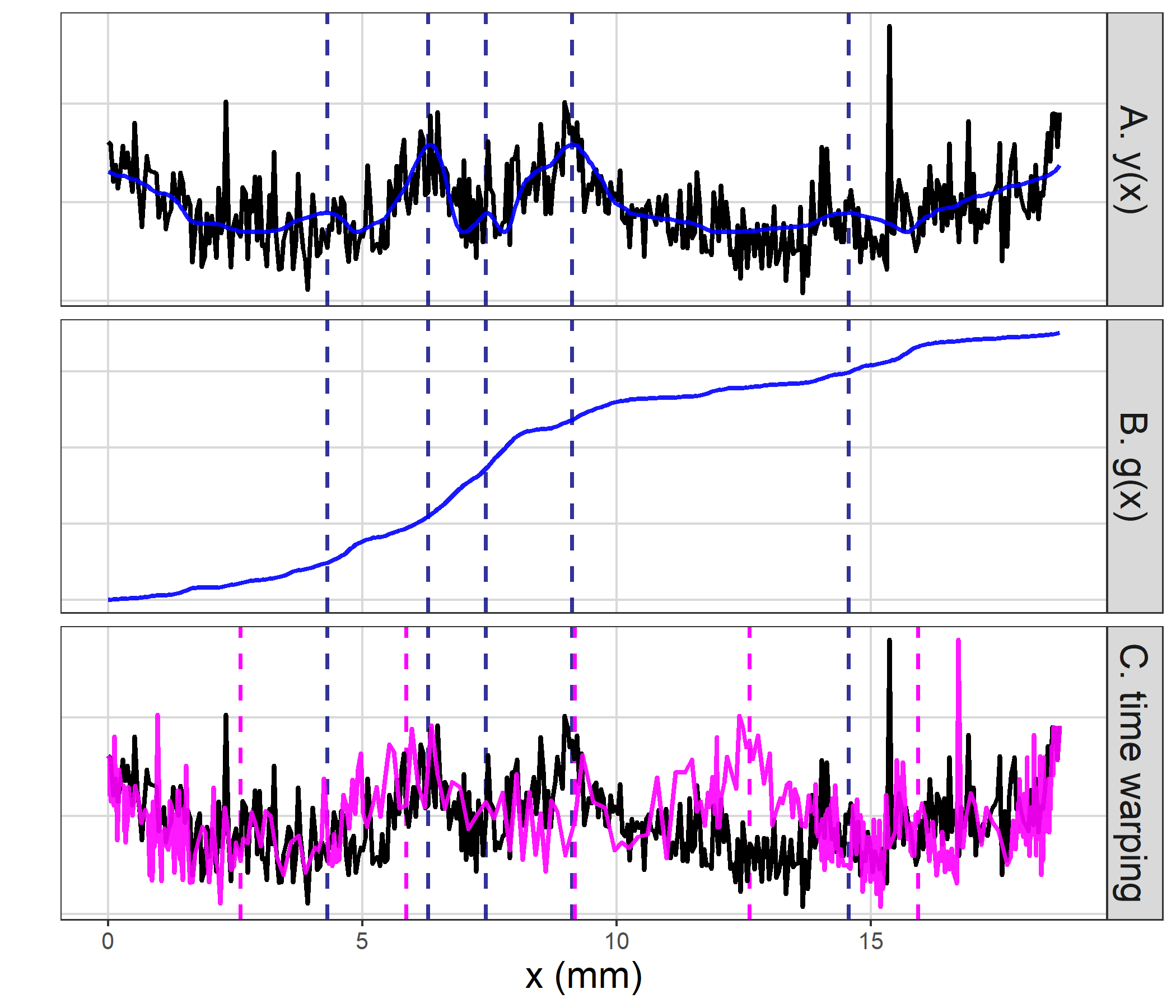}
    \end{minipage}%
    \hfill
    \begin{minipage}{0.35\linewidth}
        \caption{\textbf{Time warping of barium signal in narwhal tusk for piece 956-VII}. A. Fitted signal (blue curve) and true signal (black curve). B. Estimate $\hat{g}(x)$ of the growth process $g(x)$. C. Transformation from measured distance $x$ (black curve, same as in panel A) to estimated time $t$ (time warping, magenta curve). Blue dashed lines are estimated peak positions. Magenta dashed lines are the warped peak positions. The horizontal axis is years (lower axis) and mm (distance on tusk; upper axis).}
        \label{fig:tuskandfits}
    \end{minipage}
\end{figure}

\begin{figure}[!tb]
    \centering
    \begin{subfigure}{0.45\linewidth}
        \centering
        \includegraphics[width=\linewidth]{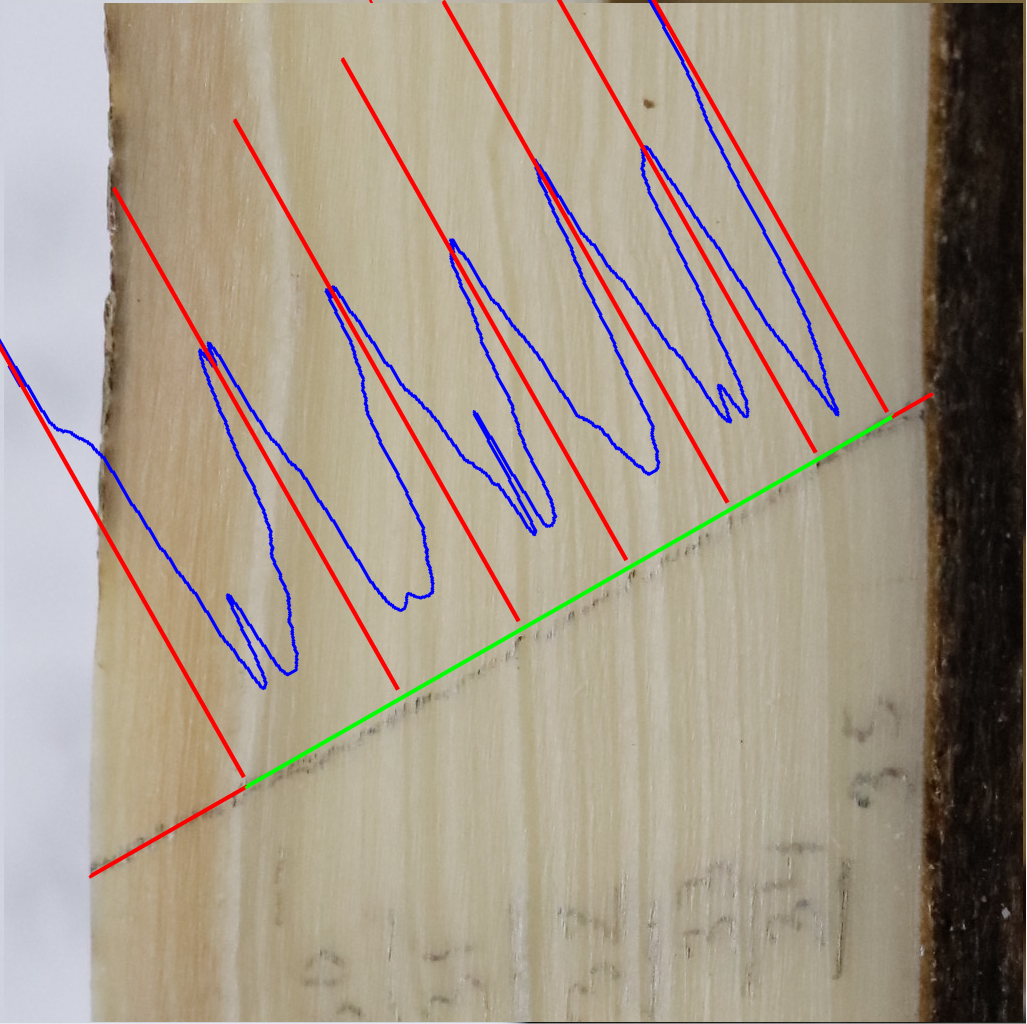}
        \caption{Tusk piece 956-IX.}
        \label{fig:subfig2}
    \end{subfigure}
    \begin{subfigure}{0.45\linewidth}
        \centering
        \includegraphics[width=\linewidth]{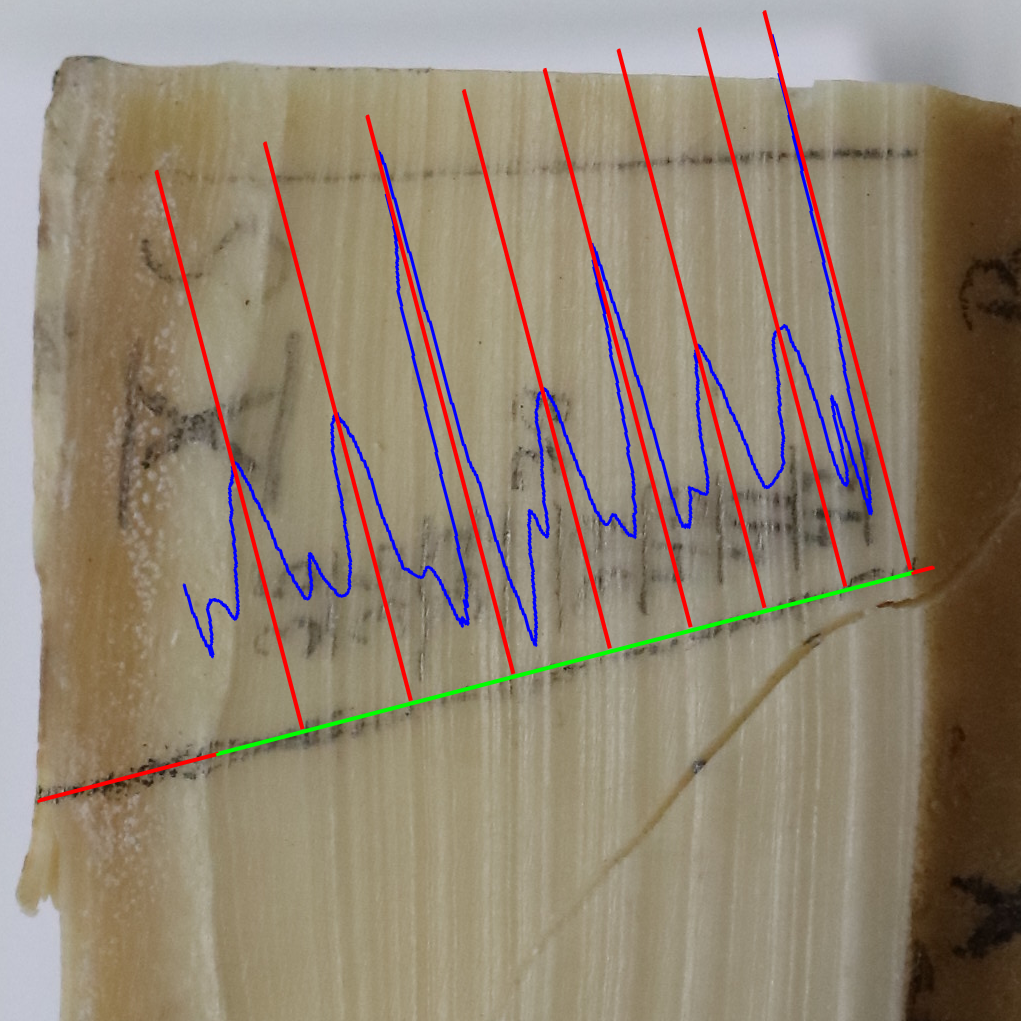}
        \caption{Tusk piece 956-Xs.}
        \label{fig:subfig3}
    \end{subfigure}
    \caption{\textbf{Fitted signals overlain tusk pieces.} Two tusk pieces with fitted signals (blue lines), estimated placement of high peaks (red vertical lines), laser line in dentin region (green segment) and pulp/cementum region (red segments).}
    \label{fig:pieces}
\end{figure}

For each piece $j$ of the tusk, we fit the model with Algorithm \ref{alg:cap} (subpanel A in Figure \ref{fig:tuskandfits} and Figure \ref{fig:pieces}). We fit each tusk piece individually allowing different parameter values in each, since the values will depend on the angle of the transect line (see red line in Figure \ref{fig2:tusk}) across the piece. For example, a steeper line will increase the length of a year, changing the distance scale, which will change all parameters related to the growth process. Likewise, amplitudes are expected to change over the lifetime of the animal, and thus, change from piece to piece.  
We obtain estimates of the growth processes $\hat{g}_j(x_{ij}), \, i = 1, \ldots , n_j, \, j = 1, \ldots , 12$, where $n_j$ is the number of observations in piece $j$ (Figure \ref{fig:tuskandfits}B). This takes us from the original coordinates $(y_{ij}, x_{ij})_{i=1,\ldots,n_j, j = 1, \ldots, 12}$ to a new set of coordinates $(y_{ij}, t_{ij})$ with $t_{ij} = \hat{g}_j(x_{ij})$, where $t_{ij}$ is an increasing sequence proportional to the number of cycles. This shift in coordinates warps the signal and adjusts for any non-linearity present in the original timeline (Figure \ref{fig:tuskandfits}C).

For each of the 12 pieces, we obtain a growth process $\hat g_j(x_{ij})$. The total number of observations is $n_{tot} = \sum_{j=1}^{12} n_j$. Re-enumerate the set of indices by concatenating chronologically
\begin{equation}
ij \rightarrow k = i+ \sum_{l=1}^j n_l, \quad k = 1, \ldots , n_{tot}.
\end{equation}
Assuming the signals can be glued together, without any gap in the timeline, we then construct an \textit{aggregated growth process}:
\begin{equation}
    \hat g(x_k) =\hat g(x_{ij}) =\begin{cases}
     \hat g_1(x_{i1}) & i \leq n_1\\
    \hat g_{j-1}(x_{n_{j-1},j-1}) + \hat g_j(x_{ij}) & n_{j-1}<i \leq n_{j}, \, j= 2, \ldots, 12
\end{cases}.
\end{equation}
 
One revolution around the unit circle corresponds to one annual cycle. Thus, the number of elapsed years at observation $m$ is $\hat g(x_m)/2 \pi$, see Eq. \eqref{eq:numberofcycles}. Since we know the time of death of the narwhal (in this case 2010), we date each observation $m$ by 
\begin{equation} 
\textrm{Year}(k) = 2010 - \left(\hat g(x_{n_{tot}}) - \hat g(x_k) \right) / 2 \pi, \quad k = 1, \ldots , n_{tot}. \label{eq: anlab} 
\end{equation}


\subsection{Age estimation} \label{ss:context}

The quantity of interest is the age of the narwhal, which is equivalent to the number of cycles easily obtained from Eq. \eqref{eq:numberofcycles} and $\hat g(x_{n_{tot}})$. A confidence interval is obtained by bootstrapping (Algorithm \ref{alg:bs}). This is accomplished by performing parametric bootstrapping on each piece, and combine the bootstrapped signals randomly to produce $M=10^7$ realizations of aggregated growth processes $\hat{g}(x_k)$, of which each has its own age estimate.
The age estimate, including upper and lower confidence bounds, are summarized in table \ref{tab:glg_estimates}. For benchmarking, the table also lists the most common age estimator - manual counting of GLGs by our expert with proficiency in handling tusks - and the age estimate of a newer, modern approach using Carbon-14 measurements \citep{Gardeetal2024}.

In the supplementary material, we present a Monte Carlo analysis to assess the coverage of our age estimates. While the nominal coverage is 95\%, our results indicate a slight undercoverage. This discrepancy is likely due to the residual bootstrap occasionally producing multi-modal distributions of the age estimates. In such cases, the algorithm may converge to a solution that corresponds to a period doubling or halving.

\begin{table}[!tb]
  \centering
  \caption{Age estimates with 95\% confidence interval. * Based on radiocarbon dating using the bomb pulse \citep{51} and Carbon-14 measurements from the narwhal tusk \citep{Gardeetal2024}.}
  \label{tab:glg_estimates}
  \begin{tabular}{ccccc}
    \toprule
    Tusk ID & Age & Age & GLGs  &  Age  \\
     & (estimate) &  (confidence interval) & (manual count) & ($^{14}C$ count)* \\
    \midrule
    956 & 54 & 52 - 57 & 57 & 54 \\
    \bottomrule
  \end{tabular}
\end{table}

\subsection{Model validation} \label{ss:modelval}

\begin{figure}[!tb]
\begin{subfigure}{.48\linewidth}
\includegraphics[width=\linewidth]{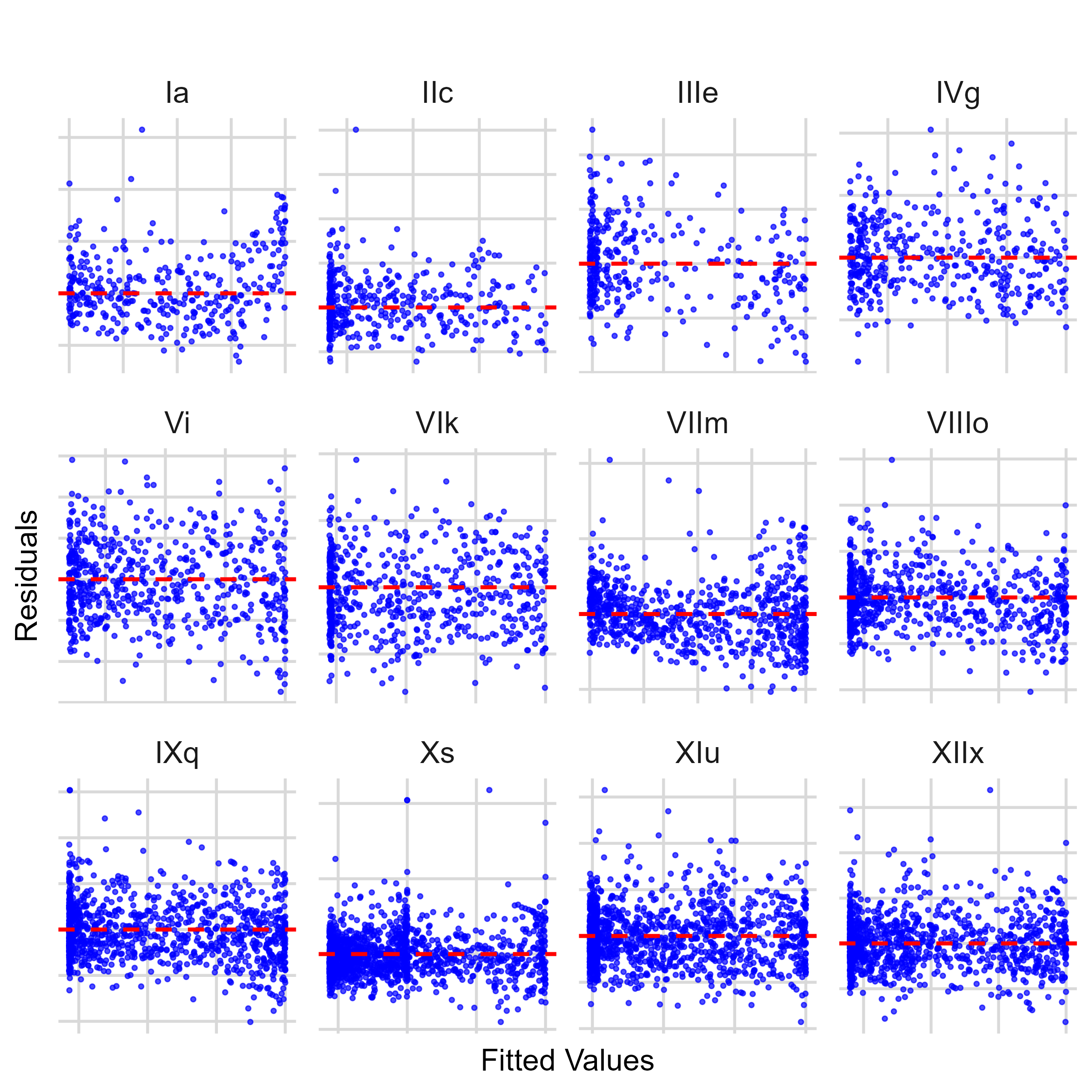}
\end{subfigure}
\hfill
\begin{subfigure}{.48\linewidth}
\includegraphics[width=\linewidth]{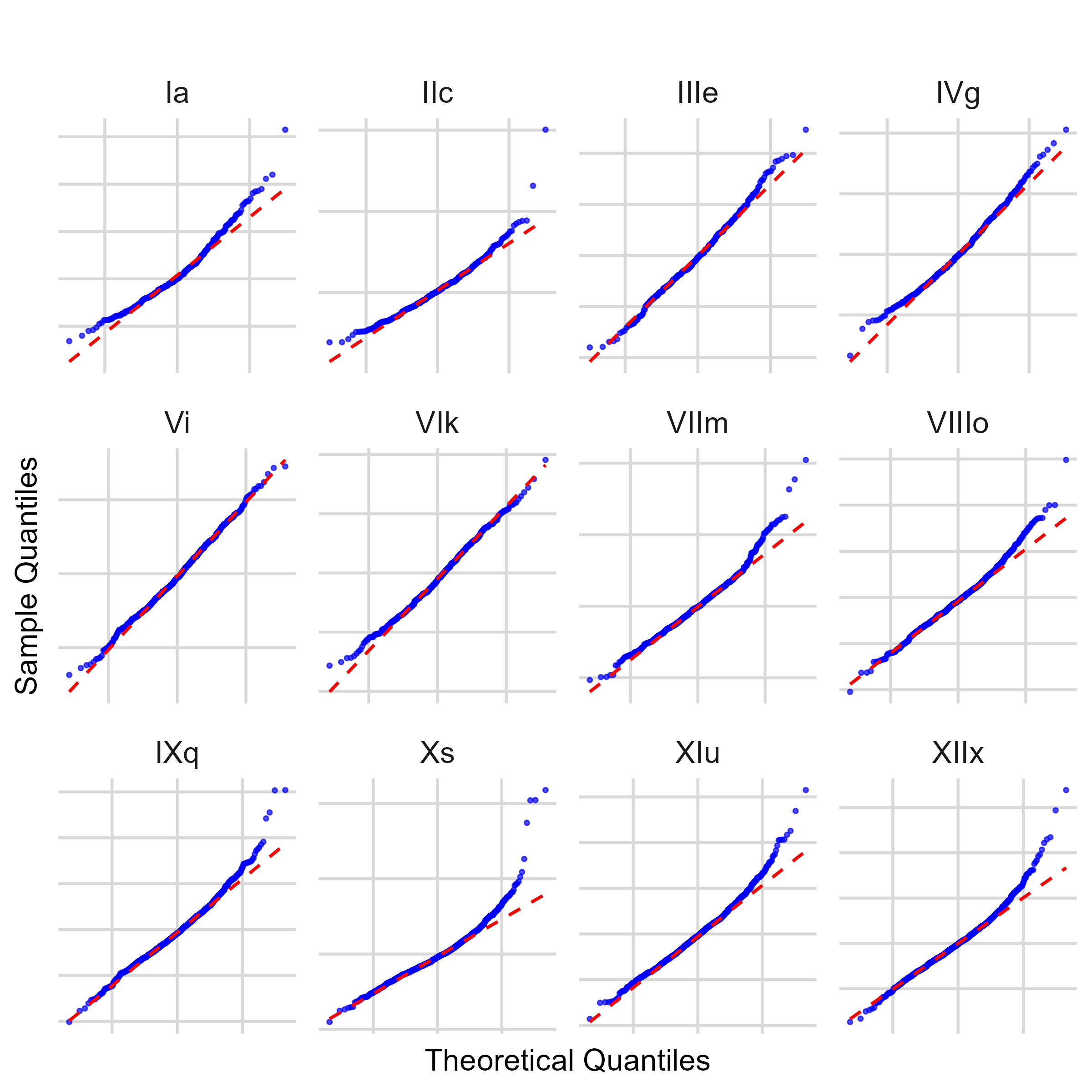}
\end{subfigure}
\caption{\textbf{Model validation plots for the case-study signals.} Residual plots (left) and QQ-plots (right) for all 12 signals of tusk 956.}
\label{fig:modelval}
\end{figure}

We have overlain the fitted curves with the tusk pieces, to compare the identified peaks with the GLGs of the tusks. In Figure \ref{fig:pieces} we show two such examples. The laser line is highlighted in green and red, when inside and outside the dentin area, respectively. The positions of the peaks are marked with vertical red lines, and we can discern how these peaks appear to align with consecutive white lines, which presumably represent one GLG. A more formal model validation was carried out by inspecting the residuals (see Section \ref{sec:residuals}) in Figure \ref{fig:modelval}. The left plot shows a residual plot (raw residuals against fitted values) for each of the 12 fitted signals and the right plot displays QQ-plots, comparing the quantiles of the raw residuals to those from the theoretical standard normal distribution. The residual plots generally conform well to the homoscedasticity assumption and show only a few outliers, likely caused by rapid secondary cycles unique to our case study signals, which are not incorporated in Eq. \eqref{eq:1}. These cycles are likely associated with sub-annual accessory layers \citep{57}. In the QQ-plots, most residuals are close to the identity line, indicating that the central tendencies are well-captured by the model. However, skewed behavior in the tails, such as in piece Xs, suggests that the model may not fully capture the behavior of extreme values. We refer to supplementary material, where we more closely explore this departure from normality for pieces XIIx, XIu, and Xs.

\begin{figure}[!tb]
\includegraphics[trim=0 0 0 12,clip,width=\linewidth,height=8cm]{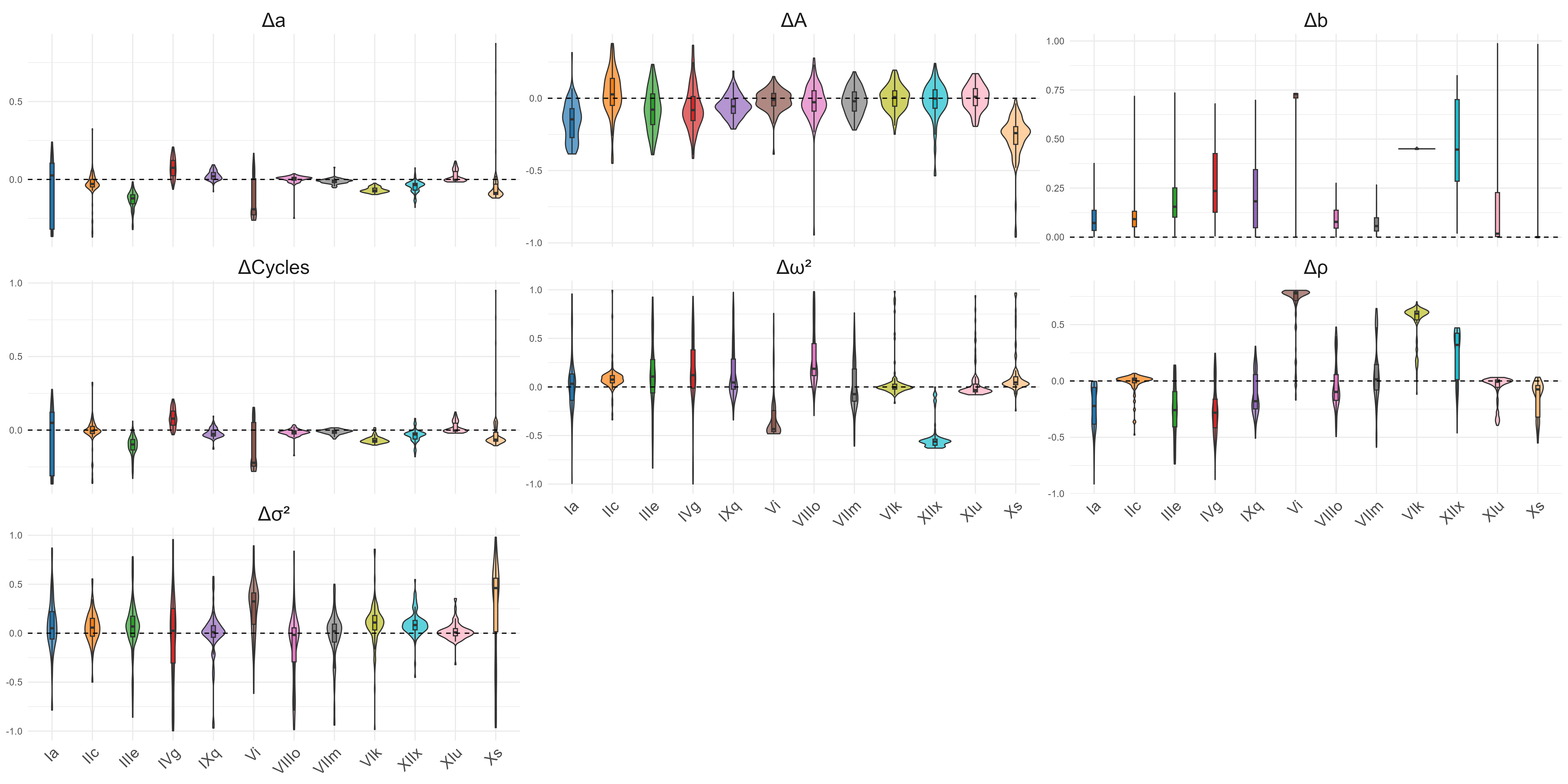}
\caption{\textbf{Estimator variability of tusk 956.} The $x$-axis represents the different tusk pieces (12 in total). For an estimate $\hat \theta$ of some model parameter $\theta$ we compute 100 bootstrap estimates $\tilde \theta$ and plot the relative differences $\Delta \theta := (\tilde  \theta - \hat \theta)/\hat \theta$. The difference in phase $\Delta b$ is computed as the shortest distance on the unit circle, not in absolute difference, thus all differences are non-negative. 
}
\label{fig:violin}
\end{figure}

Estimator uncertainties within each tusk piece is visualized in Figure \ref{fig:violin} using violinplots for all parameters, see Section \ref{sec:residuals}. 
Among the $12$ signals of the case study, several estimators display a sizeable amount of variability and bias. For example, the one-lag autocorrelation $\rho$ is highly variable and shows some unsystematic bias (i.e. bias in no particular direction) across different pieces. The infinitesimal variance $\omega^2$ is also variable, but less prone to bias. 

\section{Discussion and outlook} \label{sec:disc}
The main attraction of the model is arguably the identification of the growth process $g(x) = \int_{s=0}^{x} \xi_s \, ds$, which effectively translates into identification of the timeline of the underlying hidden process \eqref{eq:2} driving the changing periodicity. While the square-root diffusion process $(\xi_x)_{x \geq 0}$ is slightly difficult to infer, likely caused by the integration of the process, the integrated process $g(x)$ appears to be well identified. 

It is important to consider the interplay between frequency and amplitude. For any time-varying quasi-periodic signal, we need to determine whether changes are primarily driven by variations in amplitude or frequency. In this study, we assumed that instantaneous jumps in the signal reflect stochastic shifts in frequency, while amplitudes vary slowly. This lets us first estimate and fix the amplitude envelope, then normalize to constant amplitudes before fitting our model. For the narwhal tusks, these assumptions imply that rapid changes in the signal occur due to changes in growth rate (on the order of days), whereas fluctuations in concentrations of elements in the environment are slow (on the order of months or years).


While the presented model is both flexible and theoretically sound, the practical implementation still requires some choices which need a theoretical foundation, and should be tailored to the specific use. Some examples include selection of smoothing parameters in the initialization phase, runtime parameters (number of iterations inside Algorithm \ref{alg:cap}, particles in SMC and the threshold for the stopping criteria). These choices do not only assist the method in proper convergence, but are also important for controlling algorithmic runtime. These choices will serve the algorithm in producing more reliable estimates and guide the optimization procedure. We recommend to conduct preliminary simulation studies to guide the user for specific applications.

\begin{acks}[Acknowledgments]
We thank the Greenlandic hunters for providing the narwhal tusk, Jeppe Møhl for sectioning the tusk, and Tonny Bernt Thomsen and Benjamin Heredia at the Geological Survey of Denmark and Greenland for performing the elemental analyses. The tusk was imported to Denmark from Greenland on CITES import permit no. IM 0822-296/18. Finally, we would like to thank Adam Gorm Hoffman, who was working with modelling and data analysis of a narwhal tusk in the early stage of the project.
\end{acks}

\begin{funding}
This work has received funding by Novo Nordisk Foundation NNF20OC0062958;  by NordForsk (Grant number 105053), and by the French-Danish CNRS International Research Network MaDeF IRN. 
\end{funding}

\bibliographystyle{imsart-nameyear} 
\bibliography{references}       





\end{document}